\documentclass[12pt]{emulateapj}

\submitted{ApJ, in press}
\shorttitle{CO and [\ion{C}{1}] in the $\rho$ Oph Cloud} 
\shortauthors{KULESA et al.}

\begin{document}

\title{Large Scale CO and [\ion{C}{1}] emission in the $\rho$ Ophiuchi
  Molecular Cloud}  
\author{Craig A. Kulesa,
Aimee L. Hungerford\altaffilmark{1}, Christopher K. Walker}
\affil{Steward Observatory, 
University of Arizona, \\  Tucson, AZ 85721}
\email{ckulesa@as.arizona.edu, aimee@as.arizona.edu, cwalker@as.arizona.edu}
\author{Xiaolei Zhang, Adair P. Lane} \affil{Harvard-Smithsonian
  Center for Astrophysics \\ 60 Garden Street  \\ Cambridge, MA 02138} 
\email{zhang@rosette.gsfc.nasa.gov, adair@cfa.harvard.edu}

\altaffiltext{1}{Present Address: Los Alamos National Laboratory, Los
  Alamos, NM 87545} 

\begin{abstract}
We present a comprehensive study of the $\rho$ Ophiuchi molecular
cloud that addresses aspects of the physical structure and condition of the
molecular cloud and its photodissociation region (PDR) by combining
far-infrared and submillimeter-wave observations with a wide range of
angular scale and resolution.   We present 1600 square arcminute maps 
(2.3 sq. pc) with 0.1~pc resolution in submillimeter CO$(4
\rightarrow 3)$ and [\ion{C}{1}]$(^3P_1\rightarrow{^3P_0})$ line
emission from the Antarctic Submillimeter Telescope and Remote
Observatory (AST/RO), and pointed observations in the CO~$(7
\rightarrow 6)$ and [\ion{C}{1}]$(^3P_2\rightarrow{^3P_1})$ lines.
Within the large scale maps, smaller spectral line maps 
of 3000~A.U. resolution over $\sim$90
square arcminutes (0.2 sq. pc) of the cloud in CO, CS, HCO$^+$
and their rare isotopomers are made at the Heinrich Hertz 
Submillimeter Telescope
Observatory (HHT) in Arizona.  Comparison of CO, HCO$^+$, and
[\ion{C}{1}] maps with far-infrared observations of atomic and ionic
species from the Infrared Space Observatory (ISO),
far-infrared and submillimeter continuum emission, and near-infrared H$_2$
emission allows clearer determination of the physical and
chemical structure of the $\rho$ Oph photodissociation region (PDR),
since each species probes a different physical region of the cloud
structure. 
Although a homogeneous plane parallel PDR model can 
reproduce many of the observations described here, the
excitation conditions needed to produce the observed HCO$^+$ and
[\ion{O}{1}] emission imply inhomogeneous structure.  Strong chemical
gradients are observed in HCO$^+$ and CS; 
the former is ascribed to a local enhancement in the H$_2$
ionization rate, the latter is principally due to shocks.
Under the assumption of a simple two component
gas model for the cloud, we find that [\ion{C}{2}] and [\ion{C}{1}] 
emission predominantly arises from the lower density envelopes (10$^3$
to 10$^4$ cm$^{-3}$) that surround denser cloud condensations, or
``clumps''.  The distribution of [\ion{C}{1}] is very similar to
C$^{18}$O, and generally consistent with illumination from the ``far''
side of the cloud.  A notable exception is found at the the western
edge of the cloud, where UV photons create a PDR viewed ``edge-on''.
The abundance of atomic carbon is accurately modeled using a radiation
field that decreases with increasing projected distance from the
exciting star HD~147889 and a total gas column density that follows
that of C$^{18}$O, decreasing toward the edges of the cloud.  
In contrast to conclusions of other studies, we find that no
non-equilibrium chemistry is needed to enhance the [\ion{C}{1}]
abundance.   Each spectral line is traced to a particular physical 
component of the cloud \& PDR.
Though CO rotational line emission originates from both
dense condensations and diffuse envelopes, the millimeter-wave
transitions mostly find their origins in envelope material, whereas
the high-J submillimeter lines stem more from the dense clumps.
Submillimeter HCO$^+$ and infrared [\ion{O}{1}] and [\ion{C}{2}]
emission indicate clump surface temperatures of 50-200K, an
ultraviolet radiation field with I$_{\rm UV} \approx 10-90$, densities
of 10$^5$ to 10$^6$ cm$^{-3}$, and interior temperatures of $\leq
20$K.  This study highlights the value of large-scale infrared and
submillimeter mapping for the interpretation of molecular cloud
physical and chemical structure, and important future observations are
highlighted. 

\end{abstract}


\section{Introduction}

With the advent of millimeter-wave spectroscopy, large-scale surveys of
the 115~GHz CO $(J=1\rightarrow0)$ line (e.g. Dame et al. 2001) have
revolutionized our understanding of molecular clouds, 
the birthplaces of all stars.  Yet, in terms of clouds' physical
structure and evolution, many elementary questions remain shrouded in
uncertainty.   Illuminating these fundamental issues will enable 
a more complete understanding of how stars form (e.g. Bally
et al. 1987) in a Galactic context.  Sadly, numerous observational
challenges frustrate these efforts. 
For example, optically-thin spectral line diagnostics of CO that probe the
contents of the entire cloud entangle all structural cloud
components along a given line of sight.  One solution to this
problem is to constrain one's view to particular structural regions of
a cloud; for example a cloud's ``surface'', where ambient ultraviolet (UV)
photons have not been entirely absorbed by dust grains and
(e.g. H$_2$) molecules, and therefore illuminate (and dominate)
physical structure and chemical processes.

Photodissociation regions (PDR's) are these boundary interfaces,
representing the transition between
atomic and molecular phases of the interstellar medium (ISM).  Most models
invoked to explain the physics in these regions assume molecular
clouds of uniform properties surrounding the UV radiation source,
typically a neighboring early type star (Tielens
and Hollenbach 1985; van Dishoeck \& Black 1988; Burton, Hollenbach, \& 
Tielens 1990;
Hollenbach et al. 1991; Le Bourlot et al. 1993; K\"oster et al. 1994;
Kaufman et al. 1999).  UV photons with energies greater than the
hydrogen ionization threshold of 13.6 eV create an \ion{H}{2} region
surrounding the energetic star(s).  UV photons with energy less than
this threshold escape from the \ion{H}{2} region into an adjacent
molecular cloud, where photodissociation of dominant molecules like H$_2$
and CO takes place, as well as photoionization of atoms with an
ionization threshold lower than that of hydrogen -- such as carbon,
which is abundant and ionizes with photon energies as low as 11.26~eV.  
The combined effect of CO photodissociation and
C$^0$ photoionization by UV photons produces a layered structure
of C$^{+}$/C$^0$/CO surrounding the \ion{H}{2} region and extending
into the molecular cloud.  This basic picture of the C$^{+}$/C$^0$/CO
interface structure, as modeled for homogeneous cloud structures, is not in 
very good agreement with observations of
edge-on PDRs (i.e. M17); in general it is unable to account for the
extended distribution of [\ion{C}{2}] and [\ion{C}{1}] emission seen
in these regions (Stutzki et al. 1988, Keene et al. 1985).
However, models of inhomogeneous photodissociation regions have proven
capable of accounting for the extended [\ion{C}{2}] and [\ion{C}{1}]
distribution observed in some PDR's (Stutzki et al. 1988; Boiss\'e
1990; Tauber \& Goldsmith 1990;
Howe et al. 1991; Pineau Des For\^ets, Roueff, \& Flower 1992; Meixner
and Tielens 1993 [MT93]; Spaans 1996; Plume et al. 1999).  The
inhomogeneity, or ``clumpiness'', of molecular clouds is well
recognized (e.g. Blitz \& Williams 1987); it seems only
natural that such clouds' UV-illuminated surfaces should bear similar
physical traits.  

An ideal observational target for such studies is the $\rho$
Ophiuchi molecular cloud.  It is one of the nearest ($\approx$ 130 pc)
and most active 
star-forming clouds and has a rich and 
varied environment that provides the observational basis for
the current ``standard model'' of low-mass star formation (Shu 1977;
Shu, Adams, \& Lizano 1987).  The region hosts three hot B-type stars,
although 
HD~147889 appears to dominate the heating and excitation of the
cloud's atomic and molecular ``surface'' (Liseau et al. 1999).
HD~147889 is the most-embedded (E$_{\rm B-V}$=1.09; A$_{\rm V} \approx 4$)
optically-visible star that has been studied by high resolution
ultraviolet and optical spectroscopy; the line of sight passes through
the diffuse atomic and molecular ``halo'' of the $\rho$ Oph cloud
(cf. van Dishoeck \& de Zeeuw 1984; Crutcher \& Chu 1985; Cardelli \&
Wallerstein 1986).  Because of 
its deep obscuration and periphery to the main molecular cloud,
HD~147889 is thought to lie just beyond the molecular cloud, illuminating
it from behind.  Modeling of the 158 \micron\ [\ion{C}{2}] line
intensity as a function of projected distance from HD~147889 led
Liseau et al. (1999) to the conclusion that the UV-illuminated PDR
back surface of the $\rho$ Oph cloud is approximately concave.

Study of such a varied and complicated environment as $\rho$ Oph
is challenging.  High spectral and angular resolution is
required to disentangle the complicated structure of actively
star-forming environments, but simultaneously the overall
structure of the cloud only becomes apparent with studies of large
angular extent.  We have attempted to bridge these two disparate extremes by
exploiting the powers of far-infrared and submillimeter spectroscopy on
both large and small telescopes.  In \S~2 and \S~3, we present and interpret
large-scale ($40\arcmin \times 40\arcmin$) 
maps of the molecular cloud in submillimeter
CO$(J=4\rightarrow3)$ and [\ion{C}{1}] $^3P_1 \rightarrow {^3P_0}$
emission, with 25,000~A.U. (0.1 pc) resolution,
from the 1.7-meter Antarctic Submillimeter Telescope and Remote
Observatory (AST/RO).   The large maps provide a global view of the
entire cloud, enhancing the analysis of existing observations,
such as those of Liseau et al. (1999) with ISO and large scale
submillimeter continuum mapping observations such as Motte et al. (1998) and
Johnstone et al. (2000).  Regions of special interest are mapped at
3000~A.U. angular resolution in CO, CS, HCO$^+$ and their rare isotopomers
from the 10-meter Heinrich Hertz Telescope of the Submillimeter
Telescope Observatory (HHT) in Arizona.  The distribution of all of
these species, and their relevance to the structure of the $\rho$~Oph
cloud is discussed.  
In \S~4, the cloud's photodissociative properties are
explored by combining CO, HCO$^+$ and [\ion{C}{1}] data with a
thorough far-infrared spectroscopic survey of the $\rho$~Oph region using
the Infrared Space Observatory (ISO) (Liseau et al. 1999). 
We use simple one- and two-component models to compute
molecular column densities and physical conditions from
the sample of observed diagnostic atomic and molecular spectral lines.
Finally, we examine the future role of infrared H$_2$ emission as a
useful probe of the $\rho$~Oph cloud's PDR structure and excitation.

\section{Presentation of Observations}
\label{submmobs}

Several large-scale studies of the $\rho$~Oph cloud have been
performed, from millimeter-wave spectral line maps of
C$^{18}$O$(1\rightarrow0)$ pioneered by
Wilking \& Lada (1983), to infrared studies of the embedded stellar
cluster such as Barsony et al. (1997), to (sub)millimeter continuum
studies, most recently Motte, Andr\'e \& Neri (1998), Wilson et
al. (1999), and Johnstone et al. (2000).  
Liseau et al. (1999) used the Long Wavelength
Spectrometer (LWS) on ISO to
obtain continuum fluxes and integrated intensities for the
far-infrared (FIR) atomic 
fine-structure lines of [\ion{C}{2}](158~\micron) and [\ion{O}{1}](63 and
145~\micron) toward several positions of the $\rho$~Oph cloud and PDR.
To enhance the interpretation of these varied studies, we mapped
regions to maximize overlap with published data.  For example, we 
chose several east-west strip positions from Liseau et al. (1999) for
spectral line mapping.  We have also mapped over much of 
the $\rho$~Oph~A cloud core, and toward denser portions of the stellar
cluster. 

\begin{figure*}[hp]
\centering
\includegraphics[scale=1]{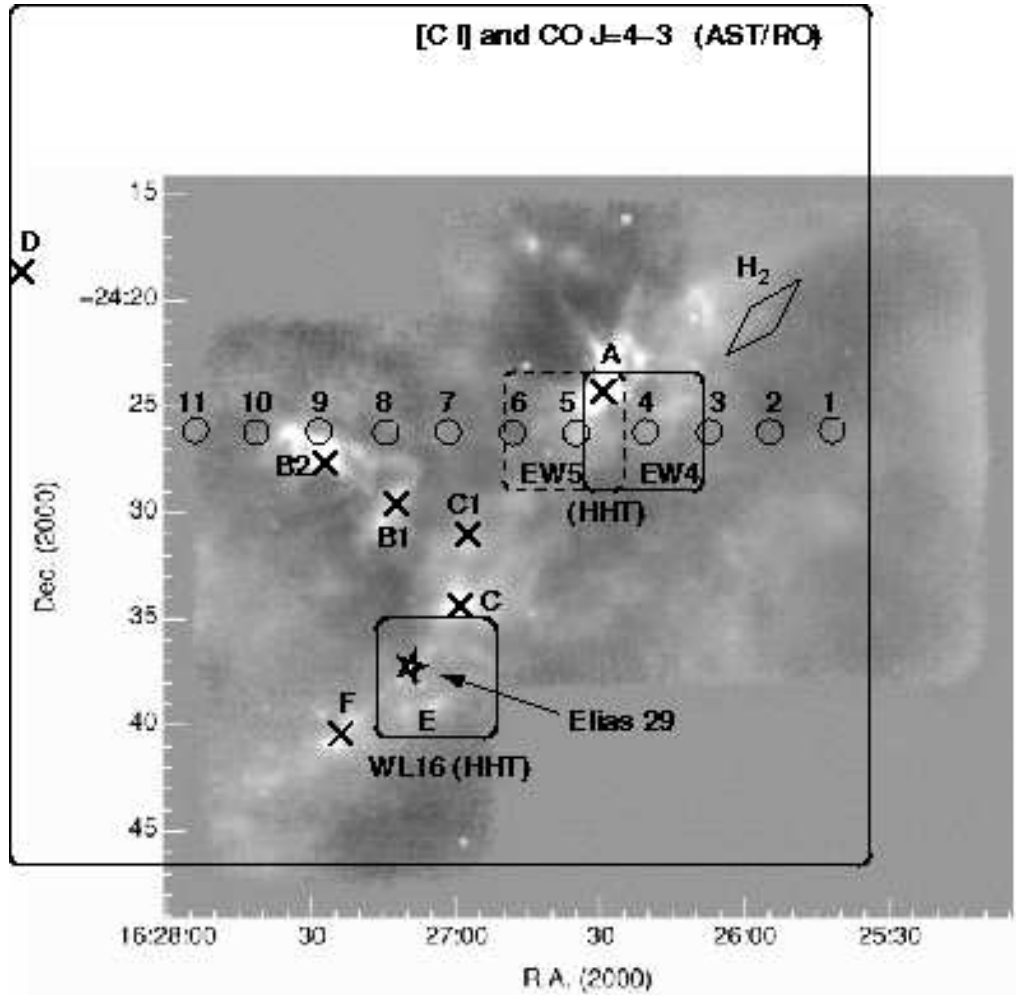}
\caption{\label{overview} Contextual overview of the main $\rho$ Oph
cloud. The grayscale image 870~\micron\ continuum map from
Johnstone et al. (2000), atop which the locations of infrared sources,
cloud cores, and the extent of our spectral line submillimeter maps
are annotated.  A region of infrared H$_2$ emission indicative of a PDR front
is labeled by a quadrilateral.}
\end{figure*}

The observations were obtained using
the AST/RO telescope at the South Pole, the Heinrich Hertz Telescope
(HHT) on Mt. Graham (Arizona), the Caltech Submillimeter Observatory
(CSO) on Mauna Kea (Hawaii), and with published data from the Infrared Space Observatory (ISO) (Liseau
et al. 1999).  These data are summarized in Table~\ref{observations}
and described in detail below. Figure~\ref{overview} provides an overview
of our submillimeter spectral line maps in the overall context of the
$\rho$~Oph cloud. 

\begin{deluxetable}{ccccc}
\tablecaption{\label{observations} Summary table of submillimeter
observations. }
\tablewidth{0pt}
\tablehead{
\colhead{Position\tablenotemark{(a)} } &
\colhead{Line} & \colhead{Telescope} &
\colhead{Map Size}& \colhead{FWHM} \\
 & & & \colhead{(\arcmin)} & \colhead{(\arcsec)}
\tablecolumns{5}
}
\startdata

WL1 & CO$(4 - 3)$&AST/RO&40$\times$40&180\\

 & CO$(7 - 6)$&AST/RO&3$\times$3&60\\

 & {[\ion{C}{1}]}$(1-0)$ &AST/RO&40$\times$40&210\\

 & {[\ion{C}{1}]}$(2-1)$ &AST/RO&3$\times$3&60\\

WL16 & CO$(2 - 1)$&CSO &5$\times$5&34\\

 & C$^{18}$O$(2 - 1)$&HHT&5$\times$5&34\\

 & C$^{18}$O$(3 - 2)$&HHT&5$\times$5&23\\

 & HCO$^+(3 - 2)$&HHT& 6$\times$5 &28\\

ISO-EW4 & C$^{18}$O$(2 - 1)$&HHT&5$\times$5&34\\

 & C$^{18}$O$(3 - 2)$&HHT&5$\times$5&23\\

 &$^{13}$CO$(2 - 1)$&HHT&5$\times$5&34\\

 & CO$(2 - 1)$&HHT&5$\times$5&33\\

 &HCO$^+(3 - 2)$&HHT&6$\times$6&28\\

 &H$^{13}$CO$^+(3 - 2)$&HHT&1.5$\times$1.5&28\\

 &HCO$^+(4 - 3)$&HHT&5$\times$5&21\\

 &CS~$(5 - 4)$ & HHT & 5$\times$5&34\\

 & CO$(7 - 6)$&AST/RO&3$\times$3&60\\

 & {[\ion{C}{1}]}$(2-1)$ &AST/RO&3$\times$3&60\\

 &{[\ion{O}{1}]}$(0-1)$ \tablenotemark{(b)} &ISO&...&62\\

 &{[\ion{O}{1}]}$(1-2)$ \tablenotemark{(b)}&ISO&...&83\\

 &{[\ion{C}{2}]}$(\frac{3}{2}-\frac{1}{2})$\tablenotemark{(b)}&ISO&...&68\\

ISO-EW5 & C$^{18}$O$(3 - 2)$&HHT&5$\times$5&23\\

 & HCO$^+(3 - 2)$&HHT&8$\times$5&28\\

 & HCO$^+(4 - 3)$&HHT&3$\times$6&28\\

\enddata
\tablenotetext{(a)}{Equatorial coordinates of central positions in
epoch J2000:
``WL1''= (16$^h$27$^m$01$^s$.5, -24$^{\rm o}$26\arcmin41\arcsec7),
WL16=(16$^h$27$^m$02$^s$.1, -24$^{\rm o}$37\arcmin25\arcsec.5),
ISO~EW4=(16$^h$26$^m$21$^s$.92, -24$^{\rm o}$25\arcmin40\arcsec.4),
ISO~EW5=(16$^h$26$^m$35$^s$.1, -24$^{\rm o}$25\arcmin40\arcsec.9)}
\tablenotetext{(b)}{Data taken from Table 2 of Liseau et al. 1999}
\end{deluxetable}

\begin{figure*}[hp]
\includegraphics[scale=0.75, angle=-90]{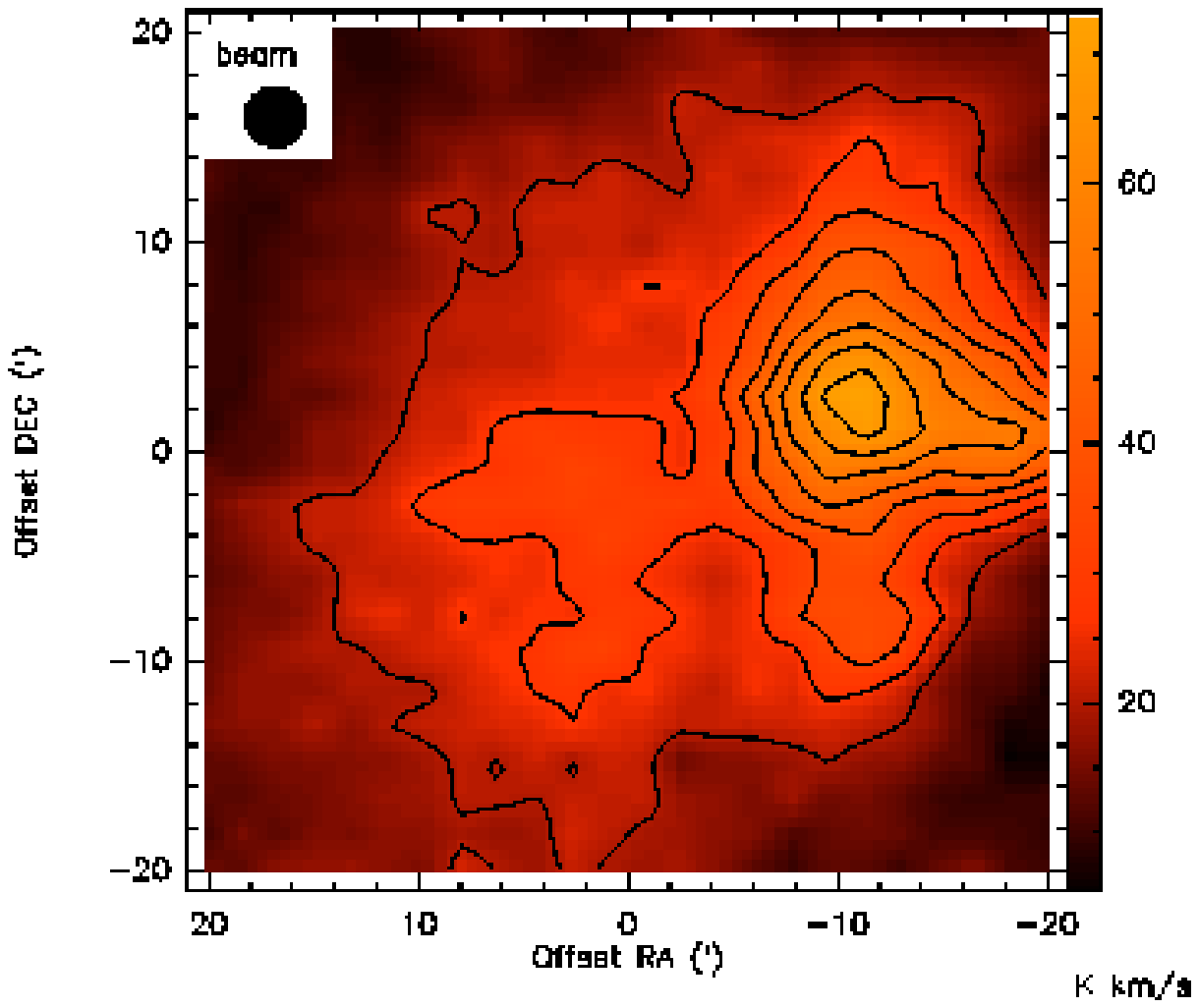}
\includegraphics[scale=0.71, angle=-90]{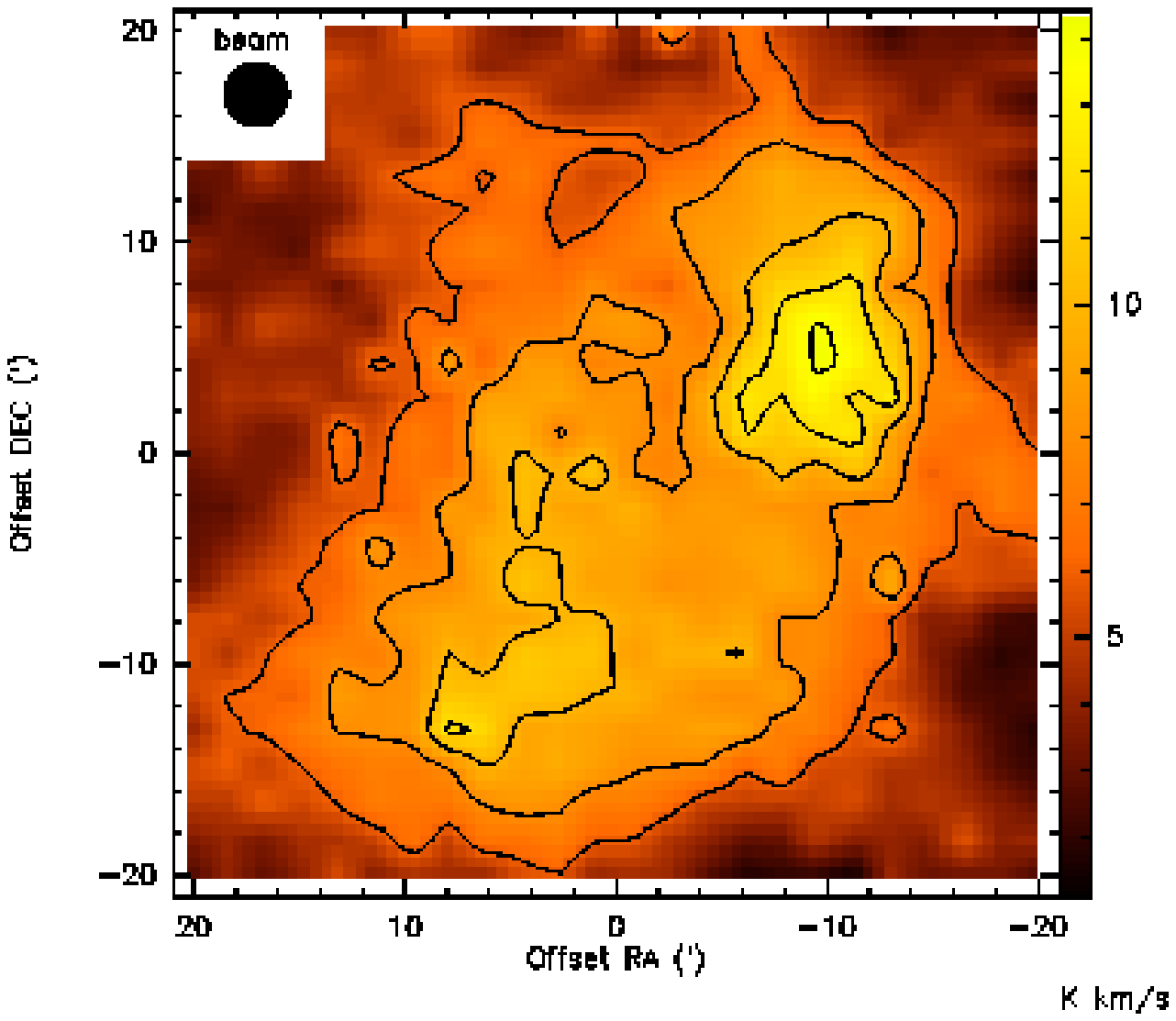}
\caption{\label{astro} ${40\arcmin \times 40\arcmin}$ map centered near
infrared source WL1 in CO$(4 \rightarrow 3)$  (a) and
[\ion{C}{1}]$(^3P_1\rightarrow{^3P_0})$ (b).  For CO$(4 \rightarrow
3)$ contours are in steps of 9$\sigma$=7~K~km~s$^{-1}$ starting at
20~K~km~s$^{-1}$.  For 
[\ion{C}{1}]$(^3P_1\rightarrow{^3P_0})$ contours are in 
steps of 4$\sigma$=2~K~km~s$^{-1}$ starting at 6~K~km~s$^{-1}$. }
\end{figure*}

Large 40\arcmin$\times$40\arcmin\ maps of the $\rho$ Oph dark cloud were made
in the 492.1607~GHz (609 \micron) fine-structure neutral carbon line and in the
461.0878 GHz (650 \micron) CO$(4 \rightarrow 3)$ line with
AST/RO in October 1998 (Figure~\ref{astro}).    Our reference
off-source position (offset +9000\arcsec\ in RA and -3000\arcsec\ in
DEC) for these observations was tested to be free of measurable CO
line emission at these frequencies. 
The maps were centered near the source WL1 (Wilking \& Lada 1983) and
oversampled (1$\arcmin$ grid in RA and DEC).  Pointed observations
were performed in the CO$(7 \rightarrow 6)$ and 809.3~GHz (370
\micron) $(^3P_2\rightarrow{^3P_1})$ 
toward the $\rho$~Oph~A cloud
core, the ISO~EW4 position of Liseau et al. (1999) and the 460/490~GHz
map center.   All AST/RO observations
were taken in position switching mode and the total integration time (on
and off) per position was 1 minute for CO$(4 \rightarrow 3)$, 3
minutes for [\ion{C}{1}] $J=1-0$ and 20 minutes for [\ion{C}{1}]
$J=2-1$.   The [\ion{C}{1}] observations were
taken with a 492~GHz quasi-optical receiver (Engargiola, Zmuidzinas \& Lo
1994, Zmuidzinas \& LeDuc 1992) in 1998 and the new 810~Ghz, 4-beam heterodyne
array receiver, {\it PoleSTAR} in 2003.  The CO$(4 \rightarrow 3)$
observations were performed in 1998 with a 
461~GHz SIS waveguide receiver (Walker, Kooi and Jacobs 2001). The
beam size at the time of the 492, 461, and 810~GHz
observations was 3.5\arcmin, 3\arcmin\ and 1\arcmin, respectively.
The line strengths were calibrated using hot and cold
loads, coupled with sky measurements at the reference position (see
Stark et al. 2001).  A
velocity resolution of 0.4~km~s$^{-1}$ was achieved for these spectra
using a 2048 channel acousto-optical spectrometer (Schieder, Tolls \&
Winnewisser 1989).  The main beam efficiency and system temperature 
for the 400~GHz receivers was
$\approx$ 0.7 and 1000-2000~K, respectively.  {\it PoleSTAR}'s
coupling efficiency, 0.5 (T$_{sys}\sim$ 20,000~K) was lower owing to
very coarse optical alignment at the time of (first light) observations.  
An instrumental velocity offset of 3 km~sec$^{-1}$ at 460 \& 490~GHz
was removed by comparison with observed CO and [\ion{C}{1}] lines
toward infrared source WL16 from the HHT and the Caltech Submillimeter
Observatory (CSO).  

\begin{figure*}[htp]
\centering
\includegraphics[scale=0.8, angle=0]{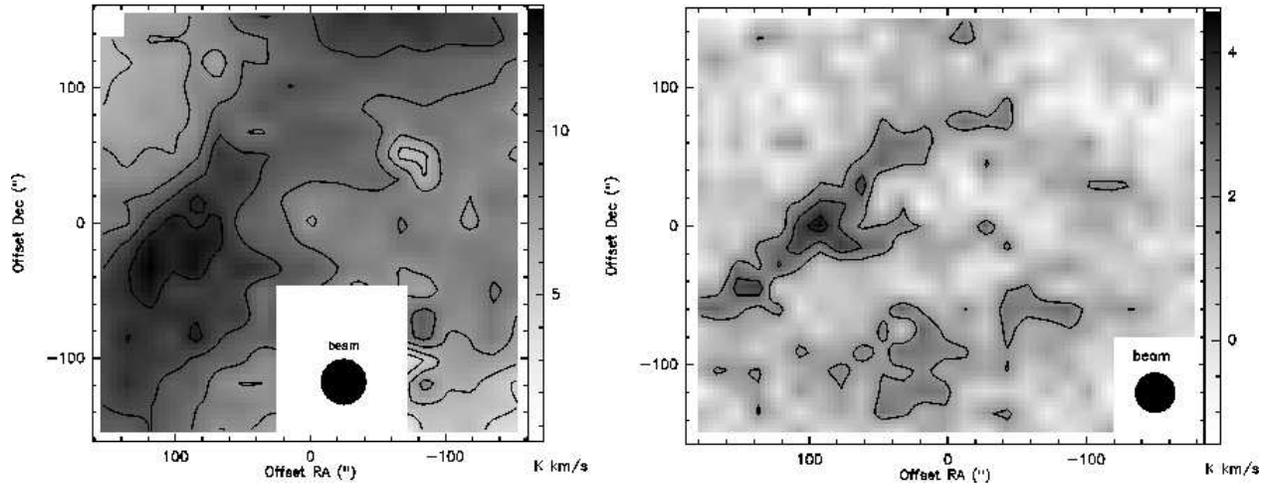}
\caption{\label{smt1} ${5\arcmin \times
5\arcmin}$ and ${6\arcmin \times 5\arcmin}$ maps from the HHT toward
WL16 in C$^{18}$O$(2 \rightarrow 1)$ and HCO$^+(3 \rightarrow 2)$.
Contours indicate 3$\sigma$=2~K~km~s$^{-1}$ starting at
3~K~km~s$^{-1}$ for C$^{18}$O$(2\rightarrow 1)$ and
1.5$\sigma$=1~K~km~s$^{-1}$ starting at 2~K~km~s$^{-1}$ for HCO$^+(3
\rightarrow 2)$. }
\end{figure*}

\begin{figure*}[htp]
\includegraphics[scale=0.55, angle=0]{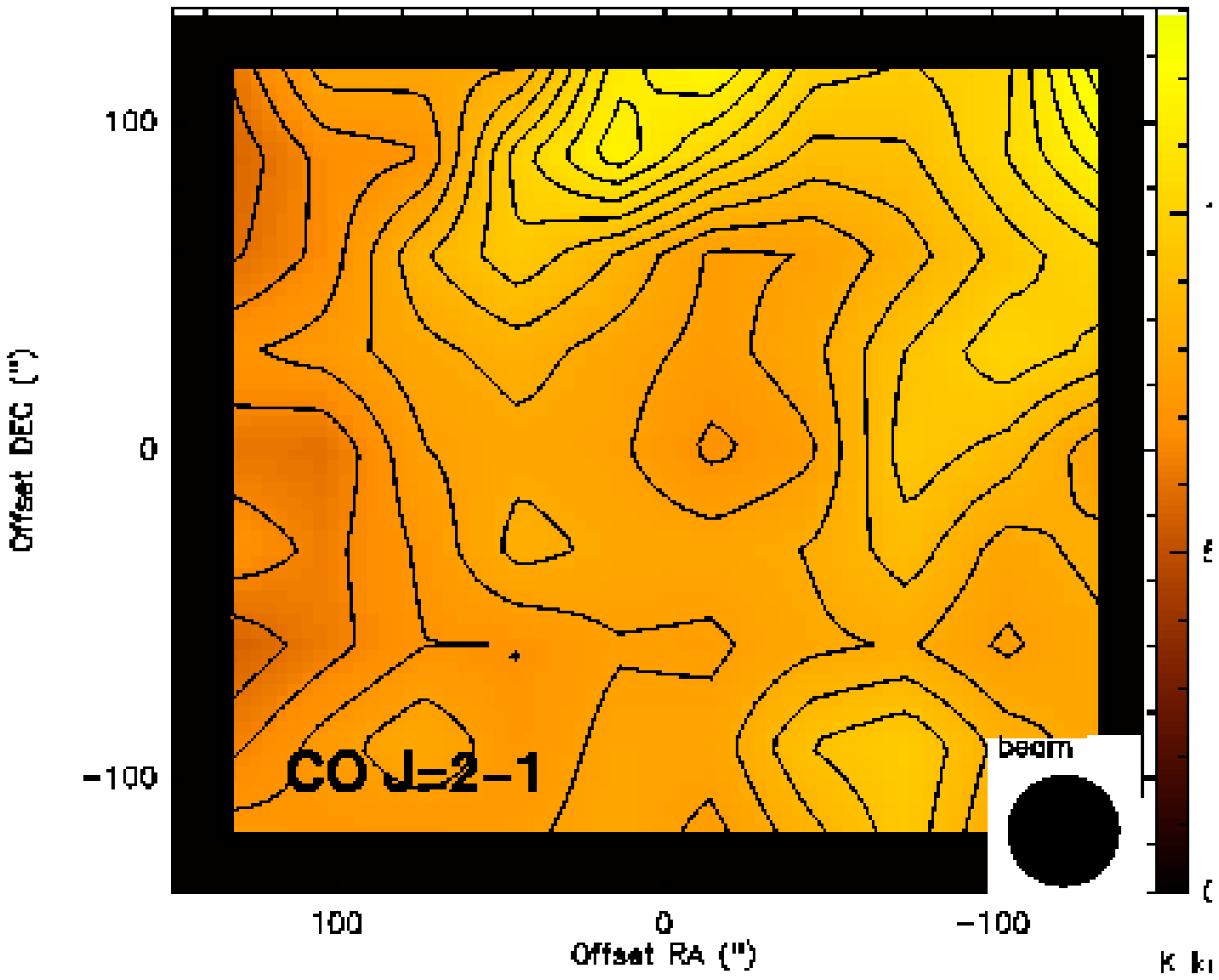}
\includegraphics[scale=0.55, angle=0]{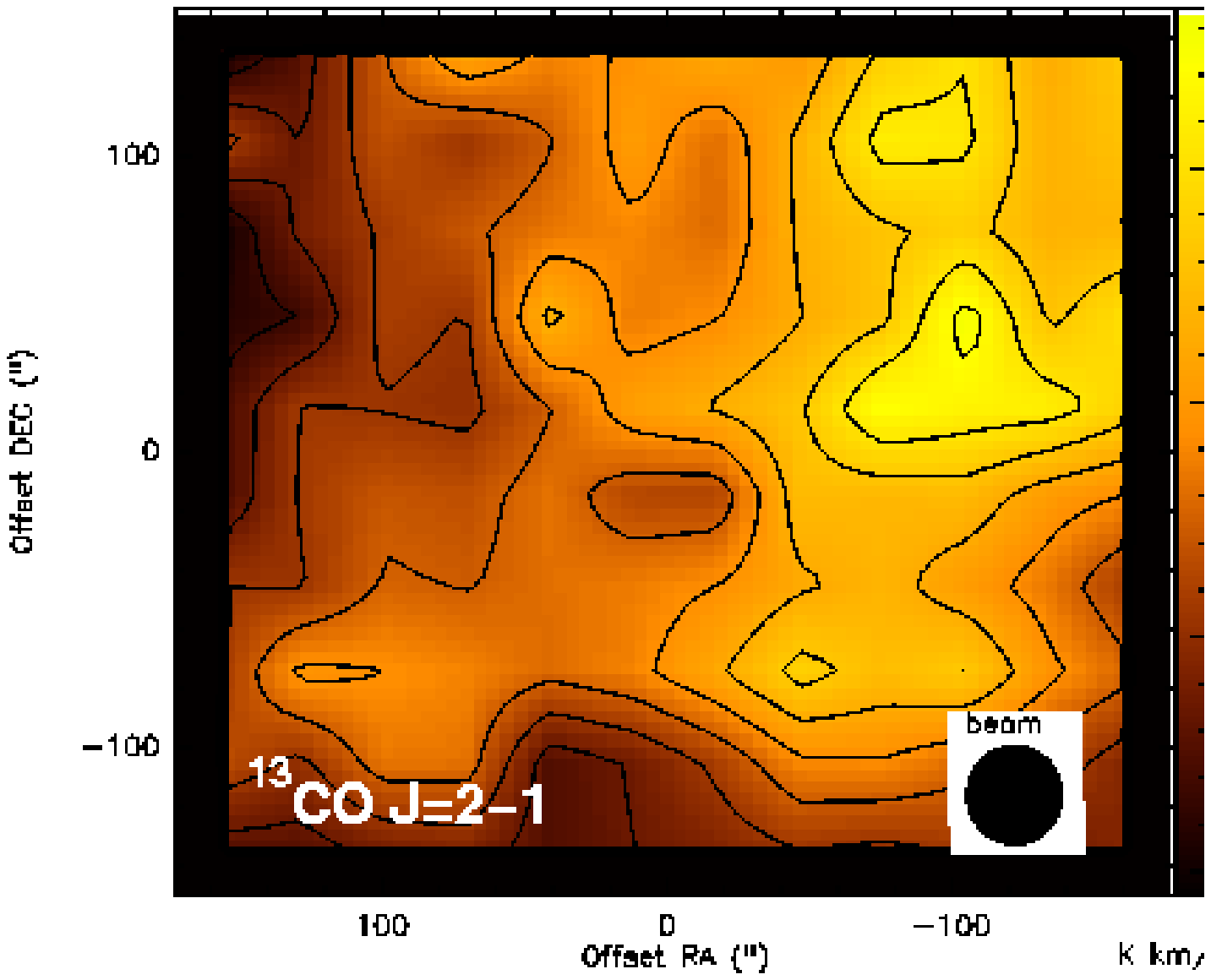}\\
\includegraphics[scale=0.55, angle=0]{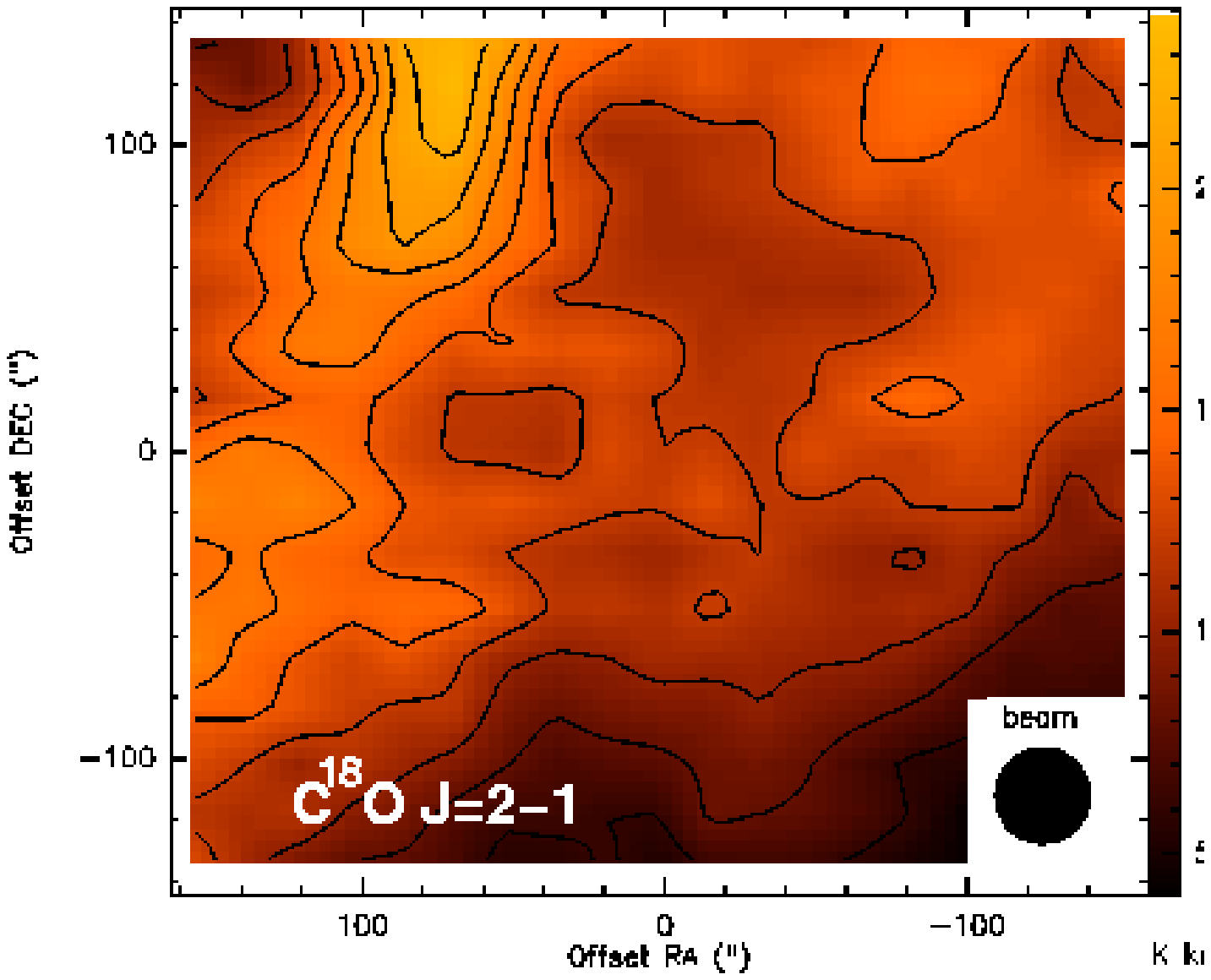}
\includegraphics[scale=0.55, angle=0]{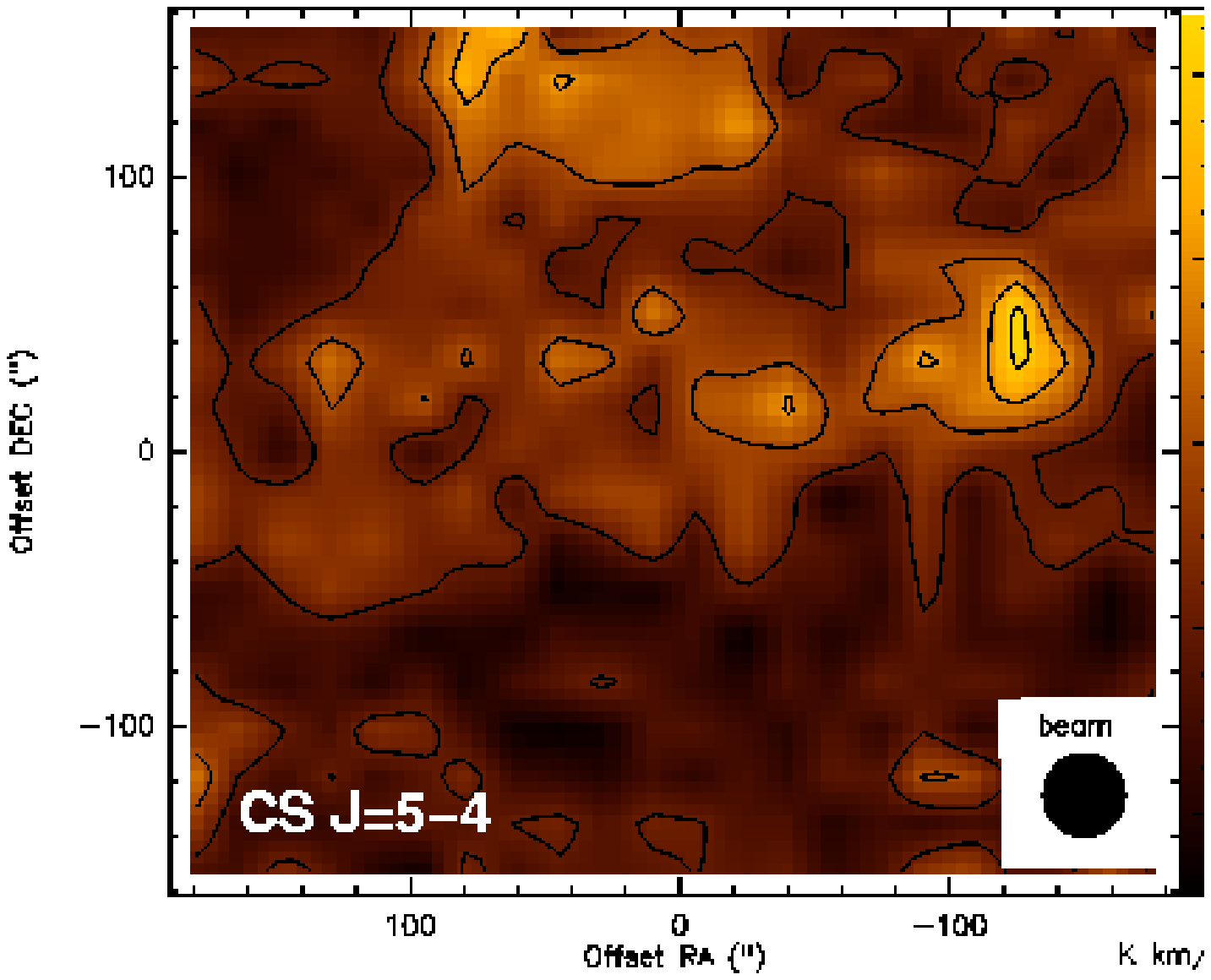}\\
\includegraphics[scale=0.55, angle=0]{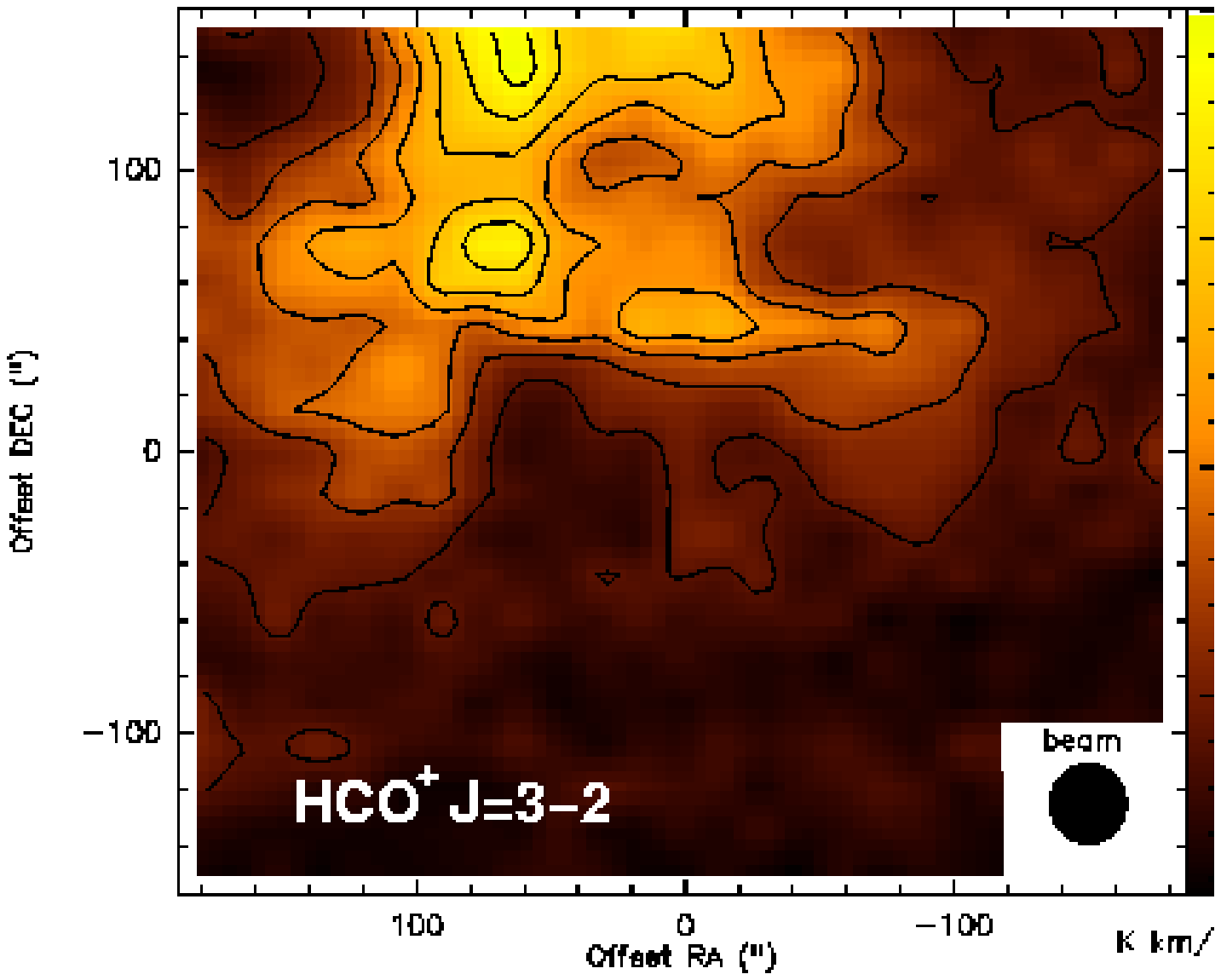}
\includegraphics[scale=0.55, angle=0]{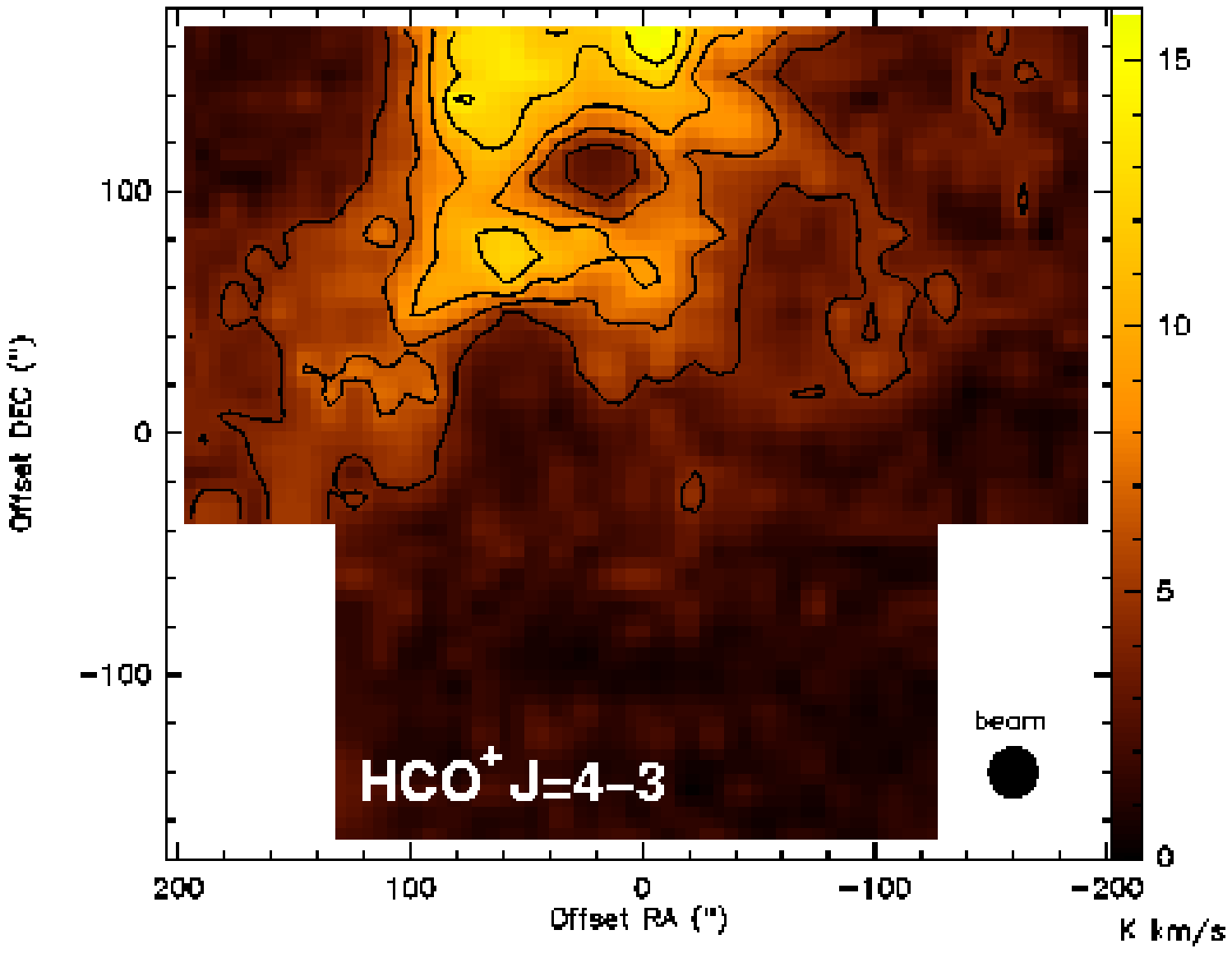}
\caption{\label{smt6} $\sim{5\arcmin \times 5\arcmin}$ maps toward
position EW4. in CO$(2 \rightarrow 1)$  (left) and $^{13}$CO$(2 \rightarrow 1)$ (right).  For
CO$(2 \rightarrow 1)$, contours are in steps of 15$\sigma$=5.5~K~km~s$^{-1}$
starting at 5.5~K~km~s$^{-1}$.  For $^{13}$CO$(2 \rightarrow 1)$, contours are in
steps of 9$\sigma$=3.9~K~km~s$^{-1}$ starting at 6.5~K~km~s$^{-1}$.  For C$^{18}$O$(2 \rightarrow 1)$ contours are in steps of
6$\sigma$=1.6~K~km~s$^{-1}$ starting at 4~K~km~s$^{-1}$.  For
CS$(5 \rightarrow 4)$ contours are in steps of
1.5$\sigma$=2~K~km~s$^{-1}$ starting at 4~K~km~s$^{-1}$. 
Contours for HCO$^+(3 \rightarrow 2)$ and HCO$^+(4 \rightarrow 3)$ are
in steps of 2.5~K~km~s$^{-1}$ starting at 4~K~km~s$^{-1}$.}
\end{figure*}

We used the HHT to obtain overlapping CO isotopomer maps (C$^{18}$O and
$^{13}$CO) for two of the east-west ISO strip positions of Liseau et
al. (1999).  Roughly
5\arcmin$\times$5\arcmin\ maps were made in C$^{18}$O$(2 \rightarrow 1)$,
C$^{18}$O$(3 \rightarrow 2)$, $^{13}$CO$(2 \rightarrow 1)$, CO$(2
\rightarrow 1)$, CS $(5\rightarrow 4)$, HCO$^+ (3\rightarrow 2)$ and
HCO$^+ (4 \rightarrow 3)$ toward the east-west strip position EW4 of Liseau et
al. (1999), and a limited subset of observations 
toward EW5 and toward infrared source
WL16 and nearby Elias~29 (Figures~\ref{smt1},\ref{smt6}). 
The maps were taken in March~1999, May~1999, and April~2001
using the spectral line
on-the-fly mapping capabilities at the HHT with facility 230~GHz
and 345~GHz SIS waveguide receivers.  A 2048~channel, 250~KHz resolution
acousto-optical spectrometer was used as a backend for both receivers,
as were three filterbanks of 1~MHz, 250~KHz, and 62.5~KHz resolution.
We adopted the same (clean) reference position
(9000\arcsec,-3000\arcsec) as used at AST/RO for the higher-frequency
observations. 
The half-power beamsize was $\sim$30\arcsec\ for the
230~GHz band observations and $\sim$20\arcsec\ for the 345~GHz band.
The adopted beam efficiencies for an extended source were
0.8 for the 230~GHz receiver and 0.7 for the 345~GHz receiver,
interpolated from measurements of point sources and the full moon. 
System temperatures with the 230~GHz receiver 
were typically 400--600~K, and 500--1500~K with the 345~GHz receiver.  
Calibration uncertainty for the AST/RO, HHT and CSO data is estimated
to be $\sim$20\% below 500~GHz and $\sim$30\% at 810~GHz.

\clearpage

\section{Morphology of the $\rho$ Oph Cloud}
\label{analysis}

\subsection{Line Excitation and Radiative Transfer}

Emergent spectral line intensities from molecular clouds are a product of the
combined effects of molecular abundances, excitation, and physical
and kinematic cloud structure.  Explicit treatment of the level
populations and transfer of line radiation is necessary to disentangle
these components and obtain a meaningful picture of the structure of
molecular clouds.
To simplest approximation, we can assume that the level populations
can be described by a Boltzmann distribution (i.e., local
thermodynamic equilibrium, or LTE), and determine total column
densities in a simplified and standard way (cf. Keene et al. 1987): 

\begin{equation}
N_T\approx{1\over g_J}exp\left( {E_J \over kT_{ex}}\right)
Q  {8 {\pi} k {\nu}^2 \over A_{_{JJ^{\prime}}}h c^3} \int T_R\ dV
\label{columneq}
\end{equation}
\noindent where $\int T_R\ dV$ is the main beam integrated intensity of the 
spectral line, $Q$ is the partition function for the energy levels
accessible under these interstellar conditions, $g_J$ is the
statistical weight of level $J$, and $A_{_{JJ^{\prime}}}$ is the
Einstein coefficient for a transition  from rotational state
$J\rightarrow{J^{\prime}}$. 
In the case of CO and its isotopomers, the populations of the lowest
rotational levels, with nominal $n_{crit} = 6 \times 10^2\ {\rm
cm}^{-3}$ in the $J=1 \rightarrow 0$ transition to $n_{crit} = 3.3
\times 10^4\ {\rm cm}^{-3}$ in the $J=4 \rightarrow 3$ transition,
should be thermalized to good approximation for an adopted density of
$n = 10^4\ {\rm cm}^{-3}$, consistent with ISO
observations of [\ion{C}{2}] and [\ion{O}{1}] line emission (Liseau et
al. 1999).  

Because the excitation of submillimeter and THz-wave lines of CO and
related atoms and molecules may be
subthermal, a more explicit calculation of the level
populations is necessary.  Assuming detailed balance and steady state,
the equations of
statistical equilibrium are solved iteratively with the level
populations and corresponding column densities, using the LTE level
populations as initial conditions.  Using the observed linewidths, 
emergent line intensities and
optical depths are computed using an escape probability formalism for
radiative transfer (de Jong, Boland, \& Dalgarno 1980). 
The competition between radiative and collisional
processes in depopulating a particular energy level typically involves a
large number of possible collisional transitions, the effects of
radiative trapping of optically thick lines, and the rates of
absorption and stimulated emission.  We adopt the collisional cross
sections from Schinke (1985), Monteiro (1985), Schr\"oder et
al. (1991), Jaquet et al. (1992), and Flower (1990) for H$_2$
interactions with CO, HCO$^+$, C$^0$, C$^+$, and O$^0$, respectively, and
suitably scaled for the appropriate kinetic temperature.  The
H$_2$ ortho-para ratio is assumed to be thermalized at the gas temperature.
Interactions of O$^0$ and C$^+$ with electrons are computed using rates from
Pequignot(1990) and Blum \& Pradhan (1992), respectively.

\begin{figure}[ht]
\centering
\includegraphics[scale=0.36,angle=-90]{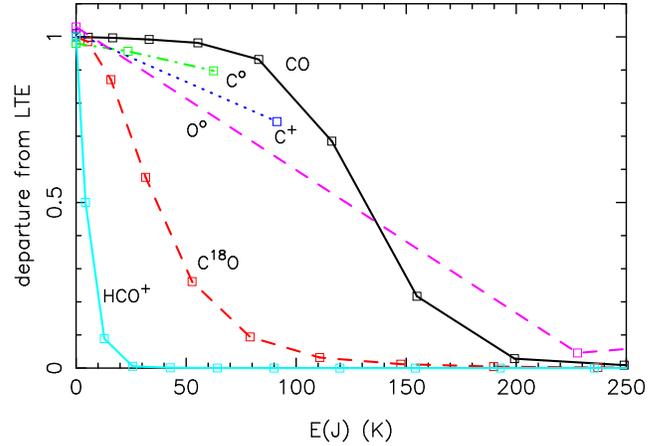}
\caption{\label{esc}  The ratio of escape probability level populations to
LTE populations as a function of energy above the ground state.  
Conditions: $n = 10^4$~cm$^{-3}$ and $T_k=25$K; column densities
scaled from N(H$_2$)=$2.5\times10^{22}\ {\rm cm}^{-2}$ with
``standard'' interstellar abundances (see text). }
\end{figure}

A sample calculation highlights the differences in excitation between
LTE and NLTE (escape probability) at moderate densities for various
PDR-probing species. 
The results are plotted in Figure~\ref{esc}.  A plane-parallel slab of
N(H$_2$)$=2.5\times 10^{22}\ {\rm cm}^{-2}$ and $n=10^4\ {\rm
cm}^{-3}$ radiates
in lines of CO, C$^{18}$O, [\ion{C}{1}], [\ion{C}{2}], and
[\ion{O}{1}].  2~km~s$^{-1}$ linewidths are adopted for the transfer
of radiation through the slab. C$^+$ and O$^0$ are subject to $T_k =
50$~K, the other species feel $T_k = 25$~K.  
The level populations are compared to the predictions
of LTE for the same {\it total} column densities as a function of energy above
the ground state.  The adopted abundances for this illustrative
example, relative to H$_2$, are [CO]~=~$2\times10^{-4}$,
[C$^{18}$O]~=~$4\times10^{-7}$, [C$^0$]~=~$2\times10^{-5}$,
[C$^+$]~=~$4\times10^{-5}$, and [O$^0$]~=~$4\times10^{-4}$.  

The submillimeter 
fine structure lines of atomic carbon are exceptionally represented by
thermal excitation, as is [\ion{C}{2}].  The very lowest rotational levels of
C$^{18}$O are thermalized, but the level populations rapidly depart
from LTE for J$\geq$5.  Radiative trapping in optically thick low-J CO
transitions maintains a thermal population distribution up to J$\leq$8, after
which it too becomes subthermally excited.  
It is useful to point out that in cold clouds with no hotter gaseous
components, the lowest rotational levels of CO are the only levels that are
populated, thusly LTE can often provide a reasonable approximation.  
Our calculation of C$^{18}$O line intensities compare
favorably with the LVG models of Mauersberger et al. (1992), which is expected
since LVG and escape probability models are functionally similar.  The
high critical 
densities of HCO$^+$ and O$^0$ lead to subthermal excitation in all
levels above the ground state.  Careful
accounting of the excitation and level populations of these species is
especially critical to the correct interpretation of their spectra.

\subsection{Morphology of Large Scale CO and [\ion{C}{1}] Emission}

With a simplified but effective mechanism for computing the NLTE level
populations 
and emergent line intensities of interstellar atoms and molecules,
we are prepared to explore the structure of the $\rho$ Oph cloud.  

\begin{figure}[!h]
\epsscale{1.15} \plotone{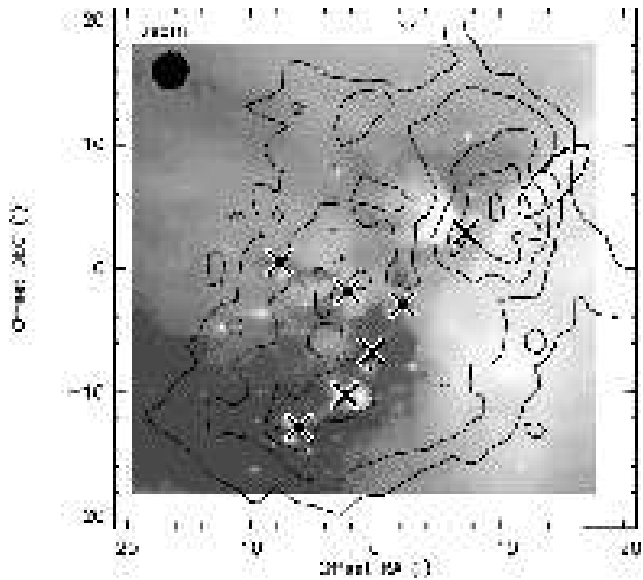}
\caption{\label{ci-iso} Overlay of the AST/RO
[\ion{C}{1}]$(^3P_1\rightarrow{^3P_0})$  map contours 
atop the Abergel et al. (1996) ISOCAM image at 7 \& 15 \micron.  The
[\ion{C}{1}] emission is extended and generally follows the edges of
warm dust emission expected for the illuminated surface of the
molecular cloud.  Crosses indicate cloud cores identified in
Figure~\ref{overview}.  The thick oval highlights a region
of intense H$_2$ and PAH emission indicative of a H$_2$
photodissociation front, from Boulanger et al. (1999). }
\end{figure}

\begin{figure}[!h]
\includegraphics[scale=0.4]{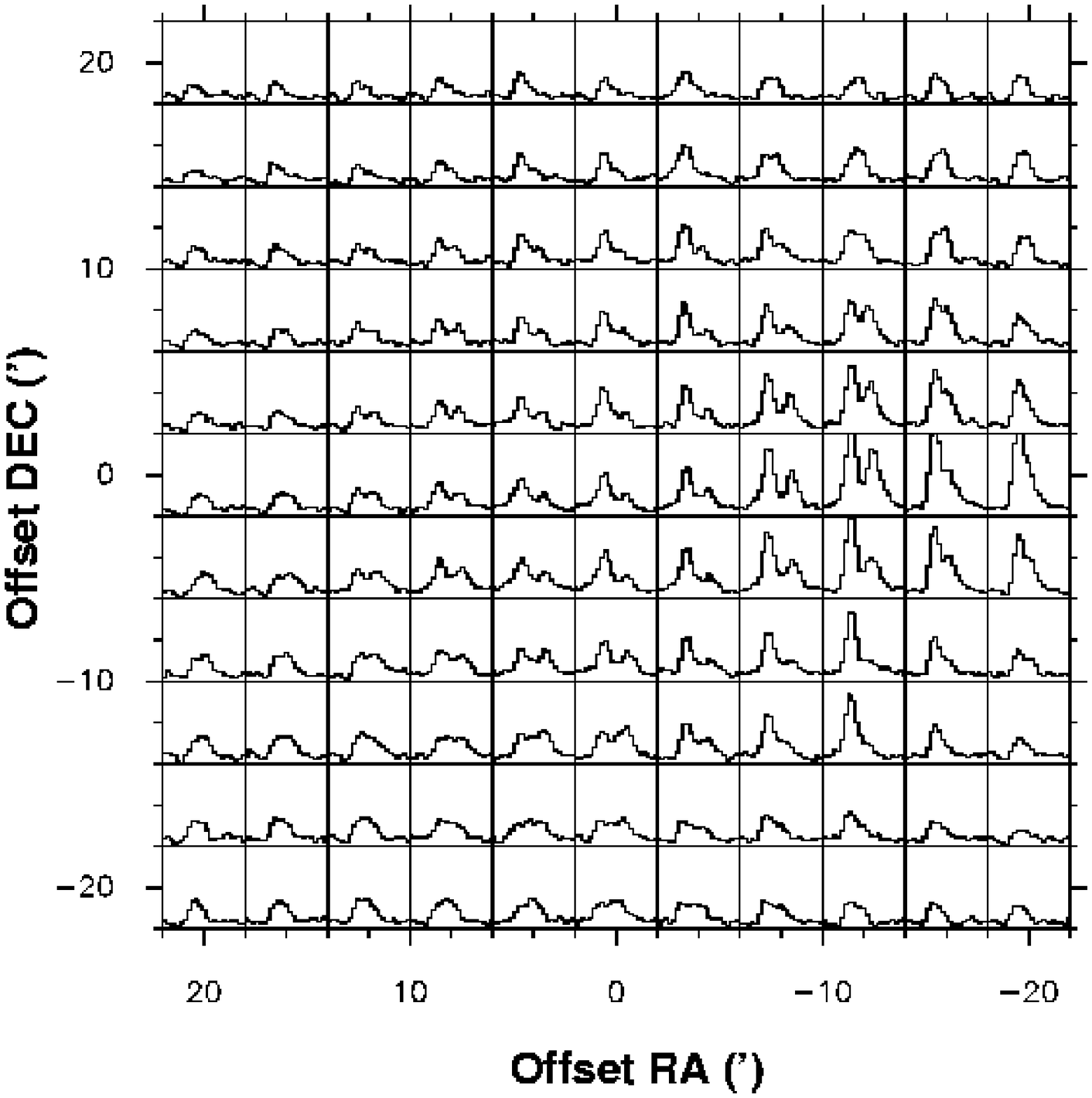}
\hfill
\includegraphics[scale=0.7]{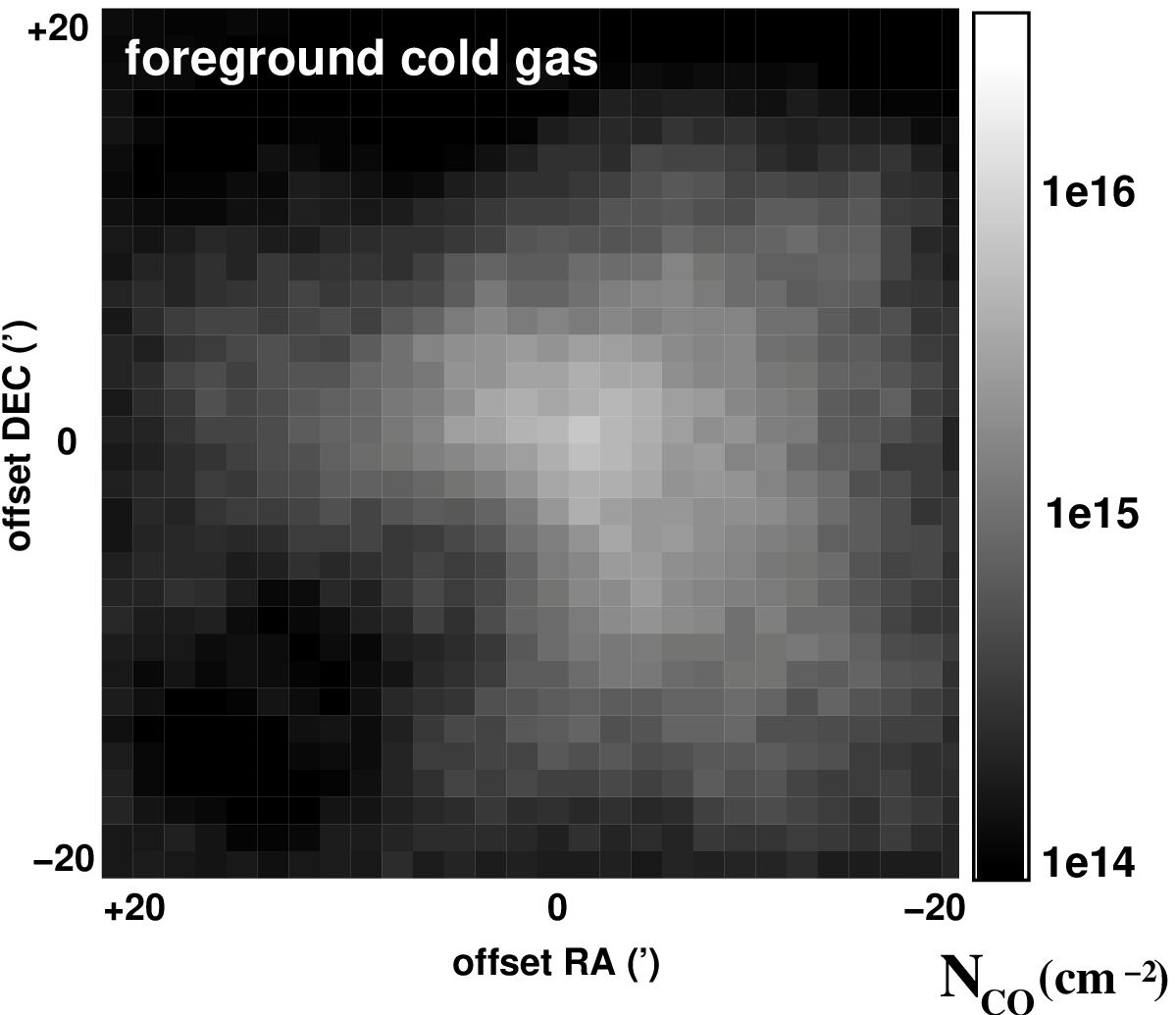}
\caption{\label{cospec} A) AST/RO spectra of optically-thick
CO$(4\rightarrow3)$ across the
$\rho$~Oph molecular cloud, resampled to a 4\arcmin\ grid for
readability.  The
velocity width for each spectrum is 9 km~s$^{-1}$, and the corrected
main beam temperature scale goes from -3 to 30K.  
Although the line profiles are severely
self-absorbed over the majority of the main cloud, Gaussian fits to
the line profiles yield estimates of
the kinetic temperature and B)  the structure of foreground absorbing CO
gas, derived from the degree of self-absorption.  
As in Figures~\ref{astro} and \ref{ci-iso}, 
offsets are relative to the map center near infrared source WL1.  }
\end{figure}

\begin{deluxetable}{ccccc}
\tablecaption{\label{excite} Thermalization of C$^0$ level populations}
\tablecolumns{3}
\tablewidth{0pt}
\tablehead{
\colhead{Position} & \colhead{$T_{MB}$(CO)} & \colhead{$T({\rm
  C^0})$}\\
\colhead{} & \colhead{(K)\tablenotemark{(a)}} & \colhead{(K)\tablenotemark{(b)}} 
}
\startdata
$\rho$ Oph A & 45 & 41 \\
WL1 & 20 & 22 \\
WL16 & 24 & 27\\
\enddata
\tablenotetext{(b)}{Derived from Gaussian fits to the peak intensity of the CO $(4
  \rightarrow 3)$ and $(7 \rightarrow 6)$ lines}
\tablenotetext{(b)}{Derived from both 809 and 492~GHz [\ion{C}{1}] lines}
\end{deluxetable}

In Figure~\ref{astro} we present the AST/RO large scale maps in
CO$(4 \rightarrow 3)$ and [\ion{C}{1}]$(^3P_1\rightarrow{^3P_0})$,
centered near infrared source WL1.  The peaks
of the two maps differ by roughly one $3\arcmin$ beam, or 0.1
pc at the distance of the $\rho$ Oph molecular cloud.  The
CO$(4 \rightarrow 3)$ line emission is very optically thick ($\tau
\approx 10^2$) where C$^{18}$O is easily observed (Wilking \& Lada
1983).  We therefore expect the peak line intensity 
to trace the kinetic gas temperature at the $\tau\sim 1$ surface
of the cloud's ``CO photosphere''.
Sample $^{12}$CO spectra across the cloud are plotted in
Figure~\ref{cospec}.  Where the CO line profile is
Gaussian, the corrected main beam antenna temperature is generally 10-15K
although it peaks at 35K at offset (-20\arcmin, 0\arcmin), consistent
with surface heating from HD~147889.  In practice, however, this analysis
is often 
complicated by prominent self-absorption in the CO line profiles toward the
cloud cores.  We fit two component (simultaneous emission \& absorption)
Gaussian functions to the $^{12}$CO line profiles and recover 
reliable peak brightness temperatures.  Since we measure absorption line
strengths as a byproduct of this analysis, we also perform a (curve of growth)
analysis of the absorption lines to yield maps of the foreground cold absorbing
gas.  Over much of the $40\arcmin \times 40\arcmin$ map, 
the excitation of the cold gas is not well
constrained, but certainly must be lower than the excitation
temperature of the line-emitting gas.  Over the range of $10 < T < 30
K$, the derived column densities vary by less than 30\% if we assume
that the cold gas is represented by $T_{ex}$=15K.  The gas is
certainly not colder than
10K, even in the unexcited eastern portion of the cloud, as the weaker
CO~$(4 \rightarrow 3)$ lines emitted there are not self-absorbed, as they
would be if the excitation of foreground gas were markedly colder than
the line-emitting gas.  

The CO emission peak, with a peak brightness temperature of $\sim$45K 
is spatially
coincident with the filamentary western boundary of the cloud.  This
strongly implies the foreground heating of the CO boundary due to UV
photoillumination and the subsequent competition for those photons by
H$_2$ molecules and dust grains.  The warm CO is localized, suggesting
that the rest of the molecular cloud's Earth-facing ``CO photosphere'' is
cold and not subject to the significant radiation field from
HD~147889 or the other B-stars attending the dark cloud. 

In contrast to $^{12}$CO, the
overall structure of the [\ion{C}{1}]$(^3P_1\rightarrow{^3P_0})$
integrated intensity map closely resembles the C$^{18}$O$(1\rightarrow0)$
integrated intensity map toward $\rho$ Oph taken by Wilking and Lada
(1983) and matches the line intensities and profiles of Frerking et al
(1989).   The [\ion{C}{1}] map morphology generally does not match the  
CO$(4\rightarrow 3)$ map, due to the large differences in line optical
depth.   The addition of the 810~GHz CO$(7 \rightarrow 6)$ and
[\ion{C}{1}]$(^3P_2\rightarrow{^3P_1})$ lines confirms that 
the [\ion{C}{1}] level populations are
thermalized at the CO excitation temperature (Table~\ref{excite}:
25-45K in the west, 20-25K 
in the map center, and 10-15K to the east).  The [\ion{C}{1}] emission
peak
has a line center optical depth of $\leq 1$, similar to
C$^{18}$O$(1\rightarrow0)$ (\S\ref{co-opacity}).  Indeed, the 492~GHz line
optical depth actually reaches a maximum of $\tau=1$ at the second
[\ion{C}{1}] peak near $\rho$~Oph~F due to the lower kinetic
temperatures expected there and the correspondingly greater populations in
the $J=0$ and $J=1$ levels. 

In Figure~\ref{ci-iso}, we overplot the same 
AST/RO [\ion{C}{1}] contours atop a gray-scale image of the $\rho$ Oph
cloud taken in 7 \& 15 \micron\ dust emission, from Abergel et al. (1996).
The warm dust probed by the mid-IR highlights the heated
``edges'' of the molecular cloud, especially the filament-like
structure at the western edge of the cloud.  This western filament is
coincident with copious rotational and ro-vibrational H$_2$ line emission
and PAH emission indicative of an H$_2$ photodissociation front
(Boulanger et al. 1999).  
The [\ion{C}{1}] emission is clearly extended, and generally follows
the edges of the warm IR dust emission, as expected for a tracer of
the warm surfaces of molecular clouds.  [\ion{C}{1}] emission peaks in
two regions; both correspond to dense condensations of dust and gas
seen in (sub)millimeter dust continuum intensity maps (Motte et
al. 1998, Johnstone et al. 2000).  The eastern intensity peak is well
correlated with the dense molecular ridge associated with the
$\rho$~Oph E and F condensations (Motte et al. 1998) and matches a
similar distribution in C$^{18}$O line emission (Wilking \& Lada
1983).  The western [\ion{C}{1}] peak lies just south-east of the
H$_2$ 
dissociation front as observed from Earth.  The C$^{18}$O peak of
$\rho$~Oph~A lies even further southeast of the [\ion{C}{1}] peak,
reminiscent of the C$^+$/C$^0$/CO arrangement of plane-parallel PDR
models.  This
suggests that some of the western PDR front is observed edge-on or at
least obliquely.  We will return to the structure of the western PDR front in
\S\ref{h2}. 

\subsubsection{Relation of [\ion{C}{1}] to [\ion{C}{2}] and CO Line Emission}
\label{kamegai}

Kamegai et al. (2003) 
also reports large scale, albeit undersampled, 
observations of the $\rho$~Oph cloud in
[\ion{C}{1}] emission, and it is of interest to compare their maps
with the present work.  The previous study interprets the
eastern [\ion{C}{1}] line emission peak (``Peak II'') as being
potentially due to the chemical youth of the region, because the
atomic carbon and C$^{18}$O integrated intensity maxima do not
follow the standard C$^+$/C/CO arrangement expected in plane-parallel PDRs.  
Indeed, an atomic-carbon-rich region could be due to the
ongoing formation of a dark cloud where the ionic, atomic, and
molecular forms of carbon have not yet equilibrated.  Although this
prospect is exciting, our study of the $\rho$~Oph cloud in both
species does not strongly support this interpretation for the
following reasons: 

\noindent {\bf 1)} Emission from atomic carbon does in fact trace the CO column
  density and dust continuum relatively faithfully.  The eastern
  ``carbon peak'' in our maps follows a northwest-to-southeast ridge
  that is also reproduced in C$^{18}$O J=1$\to$0 emission in the
  Wilking \& Lada (1983) map and also realized by our C$^{18}$O
  J=2$\to$1 and HCO$^+$ J=3$\to$2 maps around Elias 29, part of the
  atomic carbon ``ridge'' (Figure~\ref{smt6}). Furthermore, the
  $\rho$~Oph dust continuum cores ``E'' and ``F'' pass through the
  same ridge (Motte et al. 1998), including sources such as Elias~29
  and a portion of the stellar cluster (see Figures~\ref{overview} and
  \ref{ci-iso} for an overlay of our [\ion{C}{1}] map and the ISO
  infrared map with dust continuum cores labeled). 

There is some discrepancy however; the C$^{18}$O ridge continues to
the northwest, through the $\rho$~Oph continuum peak ``C'', whereas
the atomic carbon emission turns to the north instead.   Kamegai et
al. (2003) appear to use such a cut through the cloud to demonstrate
the differentiation of atomic carbon and CO.   However, they neglect
the fact that HD~147889 lies {\it beyond} the $\rho$~Oph cloud.  It is
easy to imagine that the slightly reversed [\ion{C}{1}]/CO arrangement
on the sky is entirely due to projection effects along this particular
cut.  Most of the [\ion{C}{1}] emission follows the $^{13}$CO or
C$^{18}$O distribution accurately. 

\noindent {\bf 2)} A problem of timescales exists.  The hypothesized ``carbon-rich
  peak'' in the youthful process of forming CO from C$^+$ and C$^0$
  has already been the site of active star formation for $\sim$~1 Myr,
  a typical age of young stars in the $\rho$~Oph cluster.  
  Given the high gas densities suggested
  by the abundant HCO$^+$~$(3 \rightarrow 2)$ emission cospatial with the
  [\ion{C}{1}] and CO ridge (Figure~\ref{smt6}), the chemical
  timescales for equilibrating the CO abundance are likely to be
  appreciably shorter ($<10^6$~yr) than Kamegai et al. (2003) assume.
  Their values stem 
  from the time dependent models of Lee et al. (1996), which adopt a
  maximum gas density of only 10$^4$~cm$^{-3}$, over two orders of
  magnitude less than that needed to thermalize the $J=3\rightarrow 2$
  transition of HCO$^+$.  One caveat with this interpretation lies in
  the possibility that the HCO$^+$ and atomic carbon trace different
  gas along the line of sight.  A carbon map of higher angular
  resolution will help clarify the density structure toward this
  complex region. 

\subsection{Structure of the $\rho$~Oph~A Cloud Core}

We have marked the location of our smaller scale maps taken with the
HHT in Figure~\ref{overview}.  The centers of two such maps coincide
with ISO-LWS pointings from Liseau et al. (1999) in [\ion{C}{2}] and
[\ion{O}{1}] in an east-west strip 
adjacent to the $\rho$ Oph~A cloud core.  A third mapped
region is centered on infrared source WL16 (Wilking \&
Lada 1983) and includes Elias~29.

\subsubsection{CO Line Opacity and Excitation}
\label{co-opacity}

\begin{figure}[!h]
\epsscale{1.18} \plotone{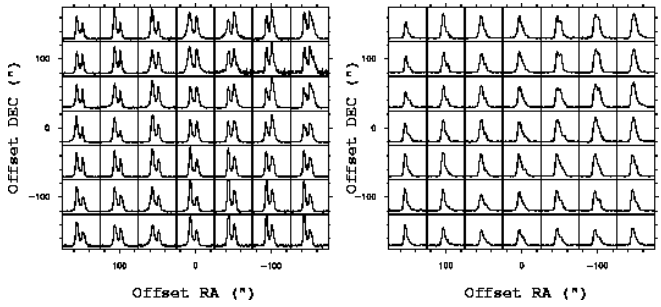}
\caption{\label{13cospec} HHT spectra of optically-thick 
CO and $^{13}$CO in the $(2\rightarrow1)$ transition toward the ISO
EW4 position; integrated intensity maps may be found in
Figure~\ref{smt6}.  Line profiles have been resampled to a 50\arcsec\
grid for readability.  As in Figure~\ref{smt6}, 
offsets are relative to the EW4 map center of Liseau et al. (1999).
For the CO map, the velocity width for is 25 km~s$^{-1}$, and the corrected
main beam temperature scale goes from -4 to 40K.  For $^{13}$CO, the
velocity width is 15~km~s$^{-1}$ per spectrum, and the corrected brightness
temperature scale goes from -4 to 30K.}
\end{figure}

\begin{figure}[ht]
\epsscale{1} \plotone{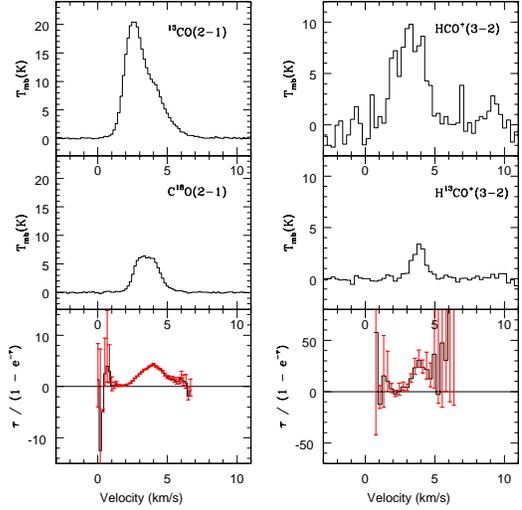}
\caption{\label{tauplot} The bottom panels portray a column density 
  correction
  for line opacity, $C(\tau) = \tau/(1-e^{-\tau})$ over observed CO
  and HCO$^+$ line
  profiles, derived from the ratioing of spectra of
  different isotopomers presented in the upper panels.
  Assumed abundances are ${\rm N(^{13}CO)}\over{\rm N(C^{18}O)}$=7.1
  and ${\rm N(HCO^+)}\over{\rm N(H^{13}CO^+)}$=70. } 
\end{figure}

Figures~\ref{smt6} depicts the HHT spectral line maps
toward ISO-EW4
which skirt the $\rho$~Oph~A cloud core and bisect a hot dust
filament seen in the ISOCAM images of Abergel et al. (1996).  
$^{12}$CO~$2\rightarrow1$ and $4\rightarrow3$ line profiles through
the region are badly contaminated by strong self-absorption
(Figure~\ref{13cospec}), indicating high optical depth and cold foreground material.
A plot of $\tau$ through the $^{13}$CO$(2 \rightarrow 1)$ line
is shown in Figure~\ref{tauplot}, where a standard 
abundance ratio of 7.1 has been adopted for
${\rm N(^{13}CO)}\over{\rm N(C^{18}O)}$.  Although this ratio is higher 
in the outermost PDR layers from poor C$^{18}$O self-shielding and subsequent selective dissociation (Black \& van Dishoeck 1988, Plume et al. 1999), 
when summed over the column of the entire cloud, this estimate is
likely to be reasonable.  We see conclusively that
the $^{13}$CO$(2 \rightarrow 1)$ emission is optically
thick with a line center optical depth of $\approx{3}$; furthermore, 
the opacity in the $(2 \rightarrow 1)$ transition of C$^{18}$O is
non-negligible, about 0.5. 
Even modest optical depths lead to radiative trapping which easily
thermalizes the J$<$3 levels for densities prevalent in dense clouds ($\geq$10$^3$~cm$^{-3}$).   The $^{13}$CO~$2\rightarrow1$ line profiles are Gaussian except toward
the $\rho$~Oph~A cloud core, toward the northeast corner of the maps
(Figure~\ref{13cospec}).  Kinetic temperatures derived throughout the region
are 20-35K, peaking at the position of the $\rho$~Oph~A2 core (Motte
et al. 1998) and along the warm dust filament that marks the western
foreground edge of the cloud. 

\begin{figure*}[htp]
\centering
\includegraphics[scale=0.66]{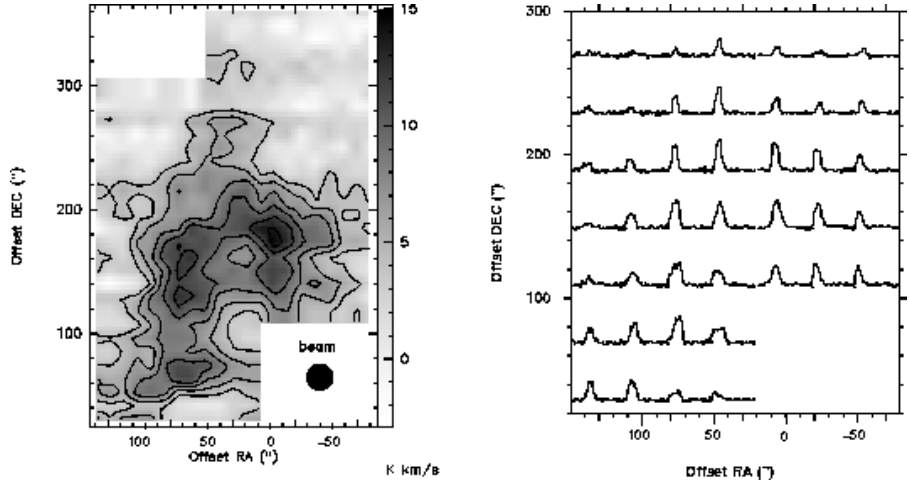} 
\caption{\label{isoew5} Extended ${3\arcmin \times 6\arcmin}$ map from the HHT
  toward the $\rho$~Oph~A cloud core (offsets from ISO EW4) in
  HCO$^+(4 \rightarrow 3)$. Contours are 
  spaced 3~K~km~s$^{-1}$ apart, starting at 6~K~km~s$^{-1}$, respectively.
The line profiles have been resampled to a 40\arcsec\ grid for
  readability.  The velocity width is 10~km~s$^{-1}$ per spectrum, and
  the corrected brightness temperature scale ranges from -4 to 15K.}
\end{figure*}

\subsubsection{Abundance and Excitation of HCO$^+$ and CS}

While examining Figure~\ref{smt6}, it is important to point out that
no two tracers of molecular material appear alike in our maps, even
those which should trace similar conditions.  In the case of CO and
its rare isotopomers, the differences result from the effects of
radiative transfer
and line optical depth, probing different depths into the cloud.
However, HCO$^+$ and CS line emission should be probing gas of high
density, yet the emission line maps share nothing in common.  Nor do
they sample the same gas seen in C$^{18}$O line emission.

For example, 
the emission in the C$^{18}$O and HCO$^+$ transitions 
strongly peak toward the NE corner of our map, coincident with the
$\rho$ Oph A cloud core and protostellar source VLA1623 (Andr\'e et
al. 1993 and references therein).  The C$^{18}$O emission in
particular is roughly comparable to the (sub)millimeter continuum emission by 
Motte, Andr\'e \& Neri (1998), Wilson et al. (1999), and Johnstone et
al. (2000).  The C$^{18}$O line emission differs principally in that the
$\rho$~Oph~A core 
dominates, and the more filamentary structures seen in the continuum
do not contrast as strongly.  For example, VLA1623 is easily separated
from the $\rho$~Oph~A core itself in the IRAM and SCUBA maps, but is
seen merely as the end of an elongated ridge of C$^{18}$O line
emission stemming from $\rho$~Oph~A and not distinguishable from it.

\begin{figure}[ht]
\centering
\includegraphics[scale=0.36,angle=-90]{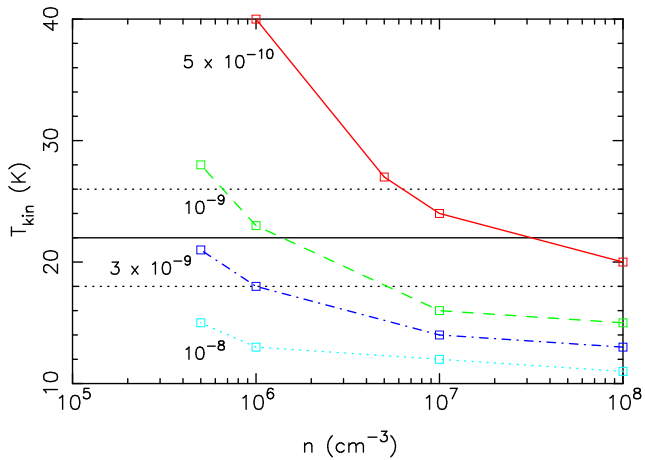}
\caption{\label{rhoophA-esc}  Physical conditions derived from the
integrated intensities of HCO$^+$~$3\rightarrow2$ and $4\rightarrow3$
in the escape-probability formalism, for the south side of the
HCO$^+$ ``ring'' (EW4 offset 10\arcsec, 60\arcsec).  Four tracks represent
different abundances of HCO$^+$, and the length of the tracks
represents the span of parameter space that fit the available
observations.  An abundance of $\leq 3 \times 10^{-9}$ best matches
the column density of H$^{13}$CO$^+$ if the isotopomer ratio is 70.
The excitation temperature of CO is plotted 
as a horizontal line, flanked by 1$\sigma$ uncertainties as dotted lines. 
}
\end{figure}

Unlike CO and its isotopomers, HCO$^+$ emission
(Figures~\ref{smt6}, \ref{isoew5}) is not prominent
throughout the map; it is instead largely confined to a ring-like
structure that includes the C$^{18}$O core of $\rho$ Oph~A at its
eastern edge.   It qualitatively reproduces the 1~mm 
submillimeter continuum
structure of Motte et al. (1998)~(figure 2a) and Johnstone et
al. (2000).  However, 
the ringed, filamentary structure is {\it more pronounced} in HCO$^+$
than it is in submillimeter continuum emission or C$^{18}$O line
emission.  It is notable that the center of the ``ring'' shows 
no sign of HCO$^+$ self-absorption, as demonstrated by the line
profile maps of Figure~\ref{isoew5}, and H$^{13}$CO$^+$ shows a
similar drop in intensity at the center of the ``ring''. 
Two possibilities explain the
heightened HCO$^+$ intensity around the $\rho$~Oph core: 
[1] a heightened HCO$^+$ abundance due to
enhanced cosmic ray flux, electron density and/or shock chemistry, [2]
a normal HCO$^+$ abundance, but in gas that is either exceptionally
warm and subthermally excited, or very dense ($n\sim 10^{6}\ {\rm cm}^{-3}$)
and thermalized near the CO excitation temperature.   

The column density and abundance of HCO$^+$ provides the best
constraints on these possibilities, and can be determined by
observations of H$^{13}$CO$^+$.  For example, a line opacity analysis
for the HCO$^+$ peak (EW4 offset +70\arcsec, +70\arcsec) that
corresponds to VLA1623 is shown in
Figure~\ref{tauplot} assuming ${\rm
N(HCO^+)}\over{\rm N(H^{13}CO^+)}$=70.  The HCO$^+(3
\rightarrow 2)$ emission is optically thick with
${\tau}\approx{25}$ at line center.   Assuming that
[HCO$^+$/H$^{13}$CO$^+$]=70, N(HCO$^+$)$=5\times10^{14}$~cm$^{-2}$.
To relate this value to that of CO and H$_2$, the column density of 
C$^{18}$O is calculated by measurement of its $3\rightarrow2$ and
$2\rightarrow1$ integrated intensities. The derived gas temperature is
25K, consistent with the observed peak antenna temperature of
$^{13}$CO, and the column density of cold C$^{18}$O is computed to be
$1.6\pm0.3 \times 10^{16}$~cm$^{-2}$, yielding N(H$_2$)$=4\pm0.9
\times 10^{22}$~cm$^{-2}$ if [C$^{18}$O]=$4\times 10^{-7}$ (Wilson \&
Rood 1994) and a strongly enhanced HCO$^+$ abundance of $1.2\times
10^{-8}$.  The same situation holds for the peak of HCO$^+$
emission at the north-west corner of the HCO$^+$ ``ring'' (see
Figure~\ref{isoew5}, offset 0\arcsec, 180\arcsec).  

The high abundance of HCO$^+$ is localized.  To illustrate this point,
the south side of the HCO$^+$ ``ring'', at
ISOEW4 offset (10\arcsec, 60\arcsec) can be seen in Figure~\ref{smt6}.  
Based on the detection of optically-thin
H$^{13}$CO$^+$~$(3\rightarrow2)$ emission of intensity 0.5~K~km~s$^{-1}$ and
an estimated H$_2$ column density of $2.5 \times 10^{22}$~cm$^{-2}$
from the $2\rightarrow1$ and $3\rightarrow2$ lines of
C$^{18}$O, the derived abundance of HCO$^+$ is $3 \times 10^{-9}$,
somewhat enhanced but not unusual.   Furthermore, the intense
submillimeter continuum peak around VLA~1623 is all but absent in
HCO$^+$ and CO emission, suggesting significant depletion, with
abundances of HCO$^+$ not even reaching 10$^{-10}$ times that of
H$_2$.  The
parameter space of kinetic temperature and density is plotted in
Figure~\ref{rhoophA-esc}.  The excitation of HCO$^+$ is consistent
with CO if it is subthermal at a density slightly less than 10$^6$~cm$^{-3}$.  
If HCO$^+$ stems from a colder environment than probed by CO line
emission, higher densities are required.

The CS molecule is often adopted as a tracer of high density
environments like HCO$^+$, yet CS $5\rightarrow4$ line emission in the
$\rho$~Oph region looks very different from HCO$^+$.  The $\rho$~Oph~A
core is nearly indistinguishable from the high column density plateau
in which it sits.  The abundance of CS at the south-west corner of the
filament, at the excitation conditions of CO and HCO$^+$ ($T_k = 25$K)
is only $7\times 10^{-10}$ relative to H$_2$ at a density of
10$^7$~cm$^{-3}$.  The derived abundances are only weakly dependent on
excitation temperature between 20 and 50K, and no CS enhancement is
seen in any of the regions where HCO$^+$ is enhanced.   On the
contrary, the CS abundance seems lowest where C$^{18}$O is
concentrated.   The CS distribution peaks 
instead at the $\rho$~Oph~A2 core (Motte et al. 1998).  The abundance, given a 
kinetic temperature of 35K from the C$^{18}$O measurements, is higher;
$2\times 10^{-9}$ at $n = 10^7$~cm$^{-3}$ to $3\times 10^{-8}$ at $n =
10^5$~cm$^{-3}$.  The CS line profile has a red shoulder, indicating that
it contributes to a (dense) outflow in the region; indeed, the line profiles of
$^{13}$CO and C$^{18}$O also indicate the presence of higher velocity
material.  The enhanced CS abundance toward $\rho$~Oph~A2 is most readily
explained by the liberation of sulphur-bearing species from dust
grains to the ISM via heating and shocks in the outflow (e.g. Mitchell
1984).  

The differences in the CS and HCO$^+$ maps suggest that
different chemical mechanisms are responsible for their relative
abundance distributions.  If shocks best explain the variation in CS, is
an enhanced ionization rate a plausible explanation for the behavior
of HCO$^+$?   To explore this possibility, we consider a simplified
set of reactions which govern the abundance of HCO$^+$, 
namely formation via the
ionization of H$_2$ to H$_3^+$ (followed by H$_3^+$ + CO $\to$ HCO$^+$
+ H$_2$), and destruction by recombination with
free electrons and reactions with gaseous water.  The latter is
negligible due to the low abundance of water in the $\rho$~Oph~A core as
measured by the Submillimeter Wave Astronomy Satellite (SWAS) (Snell
et al. 2000; Ashby et al. 2000).   We adopt an electron fraction of
$5\times 10^{-8}$, estimated coarsely from $[e] \approx (\zeta / {k_e\; n({\rm H}_2)})^{1/2}$ (Langer 1985) 
and increased by a factor of 3 to
approximately account for (a low abundance of) metal species (Lee et
al. 1996).  The ionization rate is estimated by
combining a normal cosmic ray ionization rate ($3\times
10^{-17}$~s$^{-1}$) with the local X-ray flux derived from 
the brightness and projected distance from all point sources seen in public
ACIS/Chandra images of $\rho$~Oph~A (Hamaguchi et al. 2003).  Three
sources with $L_X \sim 10^{30}$~erg~s$^{-1}$ are found within
3\arcmin\ (the magnetic B star $\rho$~Oph~S1 2.6\arcmin\ away, and two
sources within 1\arcmin, one seemingly at the position of VLA~1623).  
DoAr~21 ($L_X \sim 10^{33}$~erg~s$^{-1}$) lies at
a separation of 6.5\arcmin\ and also contributes to the total X-ray flux
(Imanishi et al. 2002).  Following Maloney et al. (1996), the 
impact of these sources upon the ionization rate of (molecular)
hydrogen is estimated to be $(2-6)\times 10^{-17}$~s$^{-1}$, comparable
to the contribution of cosmic rays.  At a density of $n_H =
10^6$~cm$^{-3}$, this enhanced ionization rate yields HCO$^+$
abundances of $10^{-8}$, matching the peaks observed toward
$\rho$~Oph~A. 

These chemical gradients make the $\rho$~Oph~A
core an interesting chemical laboratory for the study of
recent star formation and subsequent feedback upon molecular clouds
that warrants further study.

\section{PDR Modeling of the $\rho$ Oph Cloud}

\subsection{The Vicinity of the $\rho$ Oph~A Core}

The map that coincides with ISO pointing ``EW4'' (Liseau et al. 1999) slices
through the southern-most portion of the hot dust filament seen in the
Abergel et al. (1996)
ISOCAM image, at the edge of a region loosely interpreted as the
photoilluminated edge of the $\rho$ Oph cloud.  It also lies
immediately adjacent to the $\rho$ Oph~A cloud core.  Owing to the large
amount of high-quality data toward EW4, we will initially focus our
analysis toward this region.

\subsubsection{Column Densities}

\begin{figure}[h]
\centering
\includegraphics[scale=0.42]{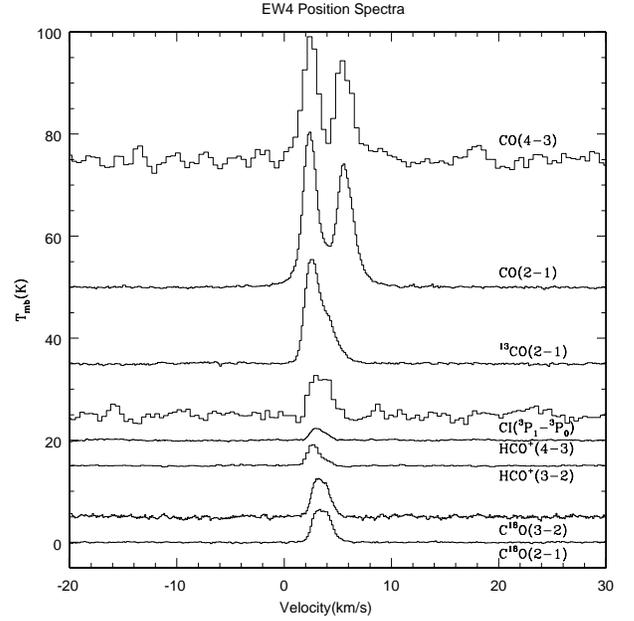}
\caption{\label{speclines} Composite ISO-EW4 spectra from the HHT
  (convolved to the AST/RO resolution of 3.5\arcmin) plotted on same
  velocity axis as the CO and [\ion{C}{1}] lines from AST/RO. }
\end{figure}

\begin{figure}[h]
\centering
\includegraphics[scale=0.36, angle=-90]{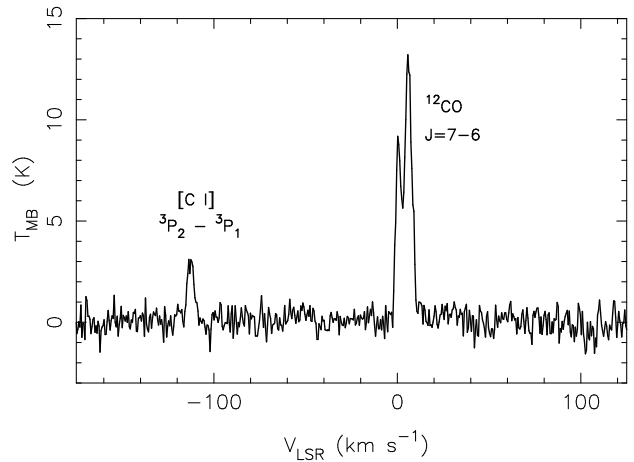}
\caption{\label{speclines2} Convolved ISO-EW4 spectra at 810~GHz of CO
 $(J=7\rightarrow6)$ and
[\ion{C}{1}] $^3P_2-^3P_1$ from the {\it PoleSTAR} array at AST/RO.}
\end{figure}

The range of telescopes used to obtain these data result in half-power 
beam sizes that vary widely between emission lines.
To compare the derived column densities in a consistent way,
we convolve our HHT spectra to the resolution of AST/RO (3.5\arcmin) and ISO
(1.1\arcmin). An indirect comparison between AST/RO and ISO
observations can then be established via the convolved HHT maps.
The convolved spectra, with $\theta_{\rm FWHM}=3.5\arcmin$, are 
plotted in Figure~\ref{speclines} along with the EW4 position spectra
of CO$(4 \rightarrow 3)$ and [\ion{C}{1}] from AST/RO.

\begin{deluxetable}{ccccc}
\tablecaption{\label{coldens} Resulting column densities toward ISO-EW4.}
\tablecolumns{5}
\tablewidth{0pt}
\tablehead{
\colhead{Position} & \colhead{Species} & \colhead{$\int T_R\ dV$}&
\colhead{$N_{tot}$} & 
\colhead{$N({\rm H_2})$} \\
\colhead{} & \colhead{} &
\colhead{$(K {km\over s})$}& \colhead{$({\rm cm}^{-2})$}
&\colhead{$({\rm cm}^{-2})$} 
}
\startdata
Offset (70\arcsec,70\arcsec) & & & & \\
 &HCO$^+$ & 22 & $5\times10^{14}$& \\
 &C$^{18}$O &26&$1.6\times10^{16}$&$4.0\times10^{22}$\\
Offset (10\arcsec,60\arcsec) & & & & \\
 &HCO$^+$ & 14 & $7.5\times10^{13}$& \\
 &C$^{18}$O &16&$1.0\times10^{16}$&$2.5\times10^{22}$\\
3.5\arcmin\ convolved beam & & & & \\
(0\arcsec, 0\arcsec) &C$^{18}$O &13.5&$7.7\times10^{15}$&$2.0\times10^{22}$\\
 &$^{13}$CO&49.1&$8.2\times10^{16}$&$2.9\times10^{22}$ \\
\enddata
\end{deluxetable}

The ratio of integrated intensities of C$^{18}$O in its $(3
\rightarrow 2)$ and $(2 \rightarrow 1)$ transitions yields
$T_{ex}~=~22$K for those levels under LTE.  An escape probability
calculation using the integrated intensities of observed C$^{18}$O and
$^{13}$CO lines yields the column densities listed in
Table~\ref{coldens} for $T_{kin}=30\ {\rm K}$ and
$n\geq\ 10^{4}\ {\rm cm}^{-3}$, implied by the antenna temperature of
$^{13}$CO~$2\rightarrow1$, 
the homogenous PDR modeling of far-infrared lines toward
ISO-EW4 by Liseau et al. (1999),
and the observed [\ion{O}{1}] and HCO$^+$ lines in the beam,
both of which have critical densities above $10^{5}\ {\rm cm}^{-3}$.
The resulting H$_2$ column density is estimated to be $2.1 \times
10^{22}$ cm$^{-2}$, which corresponds to $A_v = 26$ using standard
$A_v/E(B-V)=R=3.1$ dust 
grains, or $A_v = 18$ using $R=4.5$, likely to be more appropriate to
dense molecular environments like $\rho$~Oph (Carrasco, Strom, \&
Strom 1973).  
The derived H$_2$ column density
from $^{13}$CO is 40\% higher than that derived from
C$^{18}$O; however its higher optical depth, unmeasured excitation
temperature and non-Gaussian line profile suggests
that the C$^{18}$O observations are likely to be more reliable.
Furthermore, the escape-probability-calculated optical depths of the
C$^{18}$O lines are consistent with the previous LTE analysis that
contributed Figure~\ref{tauplot}. 

The total column densities for C$^0$, C$^+$ and O$^0$ can only be
calculated with knowledge of their excitation.    As only one line is
reliably observed for each of these species, we will now appeal to PDR
models to provide the radiation field and excitation as a function of
depth into a cloud, and will attempt to match all available observations.
One can then elucidate the types of regions from
which these three species must radiate, thereby disentangling some of
the physical structure of the $\rho$~Oph PDR.  

Liseau et al. 1999 have already performed an analysis of the [\ion{O}{1}] and
[\ion{C}{2}] emission in their ISO-LWS spectra using homogeneous PDR
models, and 
we will extend their results to include the CO, HCO$^+$ and
[\ion{C}{1}] results presented here.  We will also discuss two models
of simple inhomogeneous clouds, which are offered to provide a better
fit to all available observations.

\subsubsection{Homogeneous PDR's}
\label{homopdr}

\begin{deluxetable*}{llll}
\tablecaption{Homogeneous PDR model results for ISO-EW4}
\tablewidth{0pt}
\tablehead{
\colhead{Diagnostic} & \colhead{Observed} & \colhead{Modeled} & 
\colhead{Reference}
}
\startdata
[\ion{C}{2}] 158 \micron & $2.4 \times 10^{-4}$
\tablenotemark{(a)} & $2.5 \times 10^{-4}$ \tablenotemark{(a)}  & Kaufman
et al. (1999)\\ 

[\ion{O}{1}] 63 \micron / [\ion{C}{2}] 158 \micron & 0.43
& 0.4 & This paper\tablenotemark{(b)}\\

 &  & 0.9 & Kaufman et al. (1999)\\

[\ion{C}{1}] $^3P_1\rightarrow{^3P_0}$ & $2.3 \times 10^{-6}$ \tablenotemark{(a)} & $3.5 \times 10^{-6}$
 \tablenotemark{(a)} & Kaufman et al. (1999) \\

 & & $2.4 \times 10^{-6}$ & This paper\tablenotemark{(b)}\\

CO$(2 \rightarrow 1)$ / [\ion{C}{2}] $^3P_{3/2}\rightarrow{^3P_{1/2}}$ &
$\sim 50$ & 100 & Kaufman et al. (1999) \\

HCO$^+(4 \rightarrow 3)$ & 5.9 K$\cdot$km s$^{-1}$ & 0.36 K$\cdot$km s$^{-1}$ &
This paper\tablenotemark{(b)} \\
\enddata
\tablenotetext{(a)}{In ${\rm erg}\ {\rm s}^{-1}\ {\rm cm}^{-2}\ {\rm sr}^{-1}$}
\tablenotetext{(b)}{Cloud model follows that of van Dishoeck \& Black
(1988), see text}
\end{deluxetable*}

The archetypal PDR model invokes a plane parallel slab of constant
density, illuminated on one side 
by a source of ultraviolet photons scaled by the
local interstellar radiation field, I$_{\rm UV}$ (also G$_0$=I$_{\rm
UV} \times 2.23$), and on both sides by the cosmic microwave
background. 
The basic chemical network (including PAHs) and
heating and cooling terms (i.e. thermal structure) follows the
treatment of Kaufman al. (1999).  The PDR is gridded into zones, each
with a given
radiation field, density, and temperature.  The treatment of the
photodissociation in each zone is then
characterized by the methods described in van Dishoeck \& Black
(1988) and Black \& van Dishoeck (1987).  The rate of H$_2$
dissociation includes the effects of UV line overlap as described by
Draine \& Bertoldi (1996).  Given the 
excitation temperature derived from the 63 \micron\ [\ion{O}{1}]
transition, and based on the results of Kaufman et al. (specifically
Figure 1 of that paper),
we estimate a radiation field of I$_{\rm UV} \sim 30$ for position ISO-EW4.
We adopt a mean density of $2 \times 10^4\ {\rm cm}^{-3}$, which fits
the the 63 \micron\ 
[\ion{O}{1}] and 158 \micron\  [\ion{C}{2}] line emission as per Liseau
et al. (1999), with a slab of total column density N(H$_2$)~=~$2 \times
10^{22} {\rm cm}^{-2}$.  Our models predict the correct line intensity
for [\ion{C}{1}] $^3P_1\rightarrow{^3P_0}$ emission, $2.3 \times
10^{-6}$ erg s$^{-1}$ cm$^{-2}$ sr$^{-1}$, though the Kaufman et
al. (1999) models overestimate the [\ion{C}{1}] flux by 50\%.  
Both models similarly fit the 158 \micron\ [\ion{C}{2}] and
63 \micron\ [\ion{O}{1}] line intensity within the observational
uncertainties.  As described by Liseau et al. (1999), the 145
\micron\ [\ion{O}{1}] emission is underproduced by an order of magnitude in the
models.  Furthermore, the observed intensity of HCO$^+(3 \rightarrow 2)$ and
$(4 \rightarrow 3)$, with critical densities of $\sim2 \times 10^6$
and $\sim1 \times 10^7\ {\rm cm}^{-3}$, respectively, is not
reproducible at $n \sim 10^4\ {\rm cm}^{-3}$ unless the excitation is
exceptionally hot ($T \gtrsim 300$K) and subthermal.  Such
conditions are not indicated by the relative intensities of the
HCO$^+(3 \rightarrow 2)$ and $(4 \rightarrow 3)$ lines, which are more
consistent with a thermalized population at T$\approx 20$K, similar to
CO.  Higher densities are simply necessary to create sufficient submillimeter
HCO$^+$ emission.  A natural way to account for the discrepancy in
excitation is to adopt a heterogeneous model for the physical structure
of the region.

\subsubsection{Clumpy PDR's}
\label{clumpypdr}

\begin{deluxetable}{lll}
\tablecaption{Inhomogeneous PDR, density contrast $=30$: $3 \times
10^3\ {\rm cm}^{-3}$ \& $10^5\ {\rm cm}^{-3}$}
\tablewidth{0pt}
\tablehead{
\colhead{Diagnostic} &   
\colhead{Observed} & 
\colhead{Modeled} 
}
\tablecolumns{3}
\startdata
[\ion{C}{2}] 158 \micron & $2.4 \times 10^{-4}$
\tablenotemark{(a)} & $2.0 \times 10^{-4}$  \tablenotemark{(a)} \\ 

[\ion{O}{1}] 63 \micron / [\ion{C}{2}] 158 \micron & 0.43
& 0.36 \\

[\ion{C}{1}] $^3P_1\rightarrow{^3P_0}$ & $2.3 \times 10^{-6}$ \tablenotemark{(a)} & $2.1 \times 10^{-6}$ \tablenotemark{(a)} \\

CO$(2 \rightarrow 1)$ & $1.2 \times 10^{-6}$ \tablenotemark{(a)} & $1.0
\times 10^{-6}$ \tablenotemark{(a)}\\

HCO$^+(4 \rightarrow 3)$ & 5.9 K km s$^{-1}$ & 1.6 K km s$^{-1}$ \\
\enddata
\tablenotetext{(a)}{In ${\rm erg}\ {\rm s}^{-1}\ {\rm cm}^{-2}\ {\rm sr}^{-1}$}
\end{deluxetable}

\begin{deluxetable}{lll}
\tablecaption{Inhomogeneous PDR, density contrast $=1000$: $10^3\ {\rm cm}^{-3}$ \& $10^6\ {\rm cm}^{-3}$}
\tablewidth{0pt}
\tablehead{
\colhead{Diagnostic} &   
\colhead{Observed} & 
\colhead{Modeled}
}
\tablecolumns{3}
\startdata
[\ion{C}{2}] $^3P_{3/2}\rightarrow{^3P_{1/2}}$ & $2.4 \times 10^{-4}$
\tablenotemark{(a)} & $1.8 \times 10^{-4}$ \tablenotemark{(a)} \\

[\ion{O}{1}] 63 \micron / [\ion{C}{2}] 158 \micron & 0.43
& 1.0 \\

[\ion{C}{1}] $^3P_1\rightarrow{^3P_0}$ & $2.3 \times 10^{-6}$ \tablenotemark{(a)}  & $2.8 \times 10^{-6}$
 \tablenotemark{(a)} \\

CO$(2 \rightarrow 1)$ & $1.2 \times 10^{-6}$ \tablenotemark{(a)} & $1.0
\times 10^{-6}$ \tablenotemark{(a)}\\

HCO$^+(4 \rightarrow 3)$ & 5.9 K km s$^{-1}$ & 5.2 K km s$^{-1}$ \\
\enddata
\tablenotetext{(a)}{In ${\rm erg}\ {\rm s}^{-1}\ {\rm cm}^{-2}\ {\rm sr}^{-1}$}
\end{deluxetable}

The simplest heterogeneous models for PDR's involve the consistent 
composition and weighting of two homogeneous PDR models.  That is, we
consider a plane-parallel slab of gas containing dense condensations (clumps) of gas
randomly embedded in a more diffuse envelope, and illuminated
by an ultraviolet radiation
field.  The dense clumps individually receive an attenuated radiation field
based on the amount of intervening diffuse material photons must
travel to reach the clump. Scattering in the interclump medium is neglected. 
We consider two cases: one in which the clump:envelope
density contrast is 30:1 ($10^5$ and $3 \times 10^3$ cm$^{-3}$), and a
second in which the contrast is 1000:1 ($10^6$ and $10^3$
cm$^{-3}$). The clumps are assumed to be spherical, 0.1 parsecs in
diameter, with a volume filling factor of 0.1 and 0.2, respectively,
so as to keep the mean density comparable to the homogeneous models,
$n\approx 10^4\ {\rm cm}^{-3}$.  
The clump sizes are consistent with the assumed
sizes for clumps in other multicomponent PDR models (e.g. Meixner \&
Tielens 1993; Spaans 1996) and correspond to the angular size of
major C$^{18}$O and HCO$^+$ condensations in our spectral line maps. 

\begin{figure}[!ht]
\epsscale{1.1} \plotone{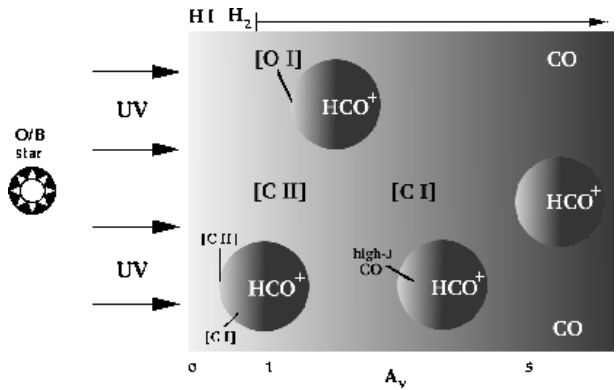}
\caption{\label{clumpfig} Illustration of a clumpy PDR and the
physical region each emission line probes.}
\end{figure}

The two inhomogenous PDR models give similar qualitative results, which are
summarized in Figure~\ref{clumpfig}.  The large majority of the [\ion{C}{1}]
$^3P_1\rightarrow{^3P_0}$ emission comes from the diffuse envelopes, 
as has been suggested by Spaans (1996) and MT93, among
others.  The envelope exceeds the low critical density and
excitation energy needed to produce significant [\ion{C}{1}] line
emission, and as it dominates the filling factor of
the region, it also dominates the production of [\ion{C}{1}]
$^3P_1\rightarrow{^3P_0}$.  This is especially true of the
high-density-contrast model.  The excitation of the [\ion{C}{1}] is
expected to be similar to that of $^{13}$CO and C$^{18}$O, which is
borne out by the similar distribution of [\ion{C}{1}] and C$^{18}$O in
$\rho$ Oph, as well as the similarities in line opacity and profile.  

The somewhat higher excitation temperatures and critical densities of
[\ion{C}{2}] emission make its point of origin slightly more
ambiguous.  Almost 2/3 of the [\ion{C}{2}] emission still
arises in the more diffuse envelopes, but the remaining line emission
stems from the warm surface layers of the denser clumps, particularly
those closest to the PDR front.  

Low-J $^{12}$CO line emission is highly saturated, and hence comes
from the $\tau \sim 1$ surface layer of the molecular cloud
``photosphere'' at the frequency of each CO line, and is equally excited in
the diffuse envelope and the surfaces of the clumps that are near the
PDR ``surface''.  Since the emergent line intensity is 
weighted by filling factor, most low-J line emission comes from the
diffuse envelope.   As one moves into the submillimeter and
far infrared bands, the less-saturated high-J lines arise most
prominently from the warm, dense clump surfaces.  Line emission from
the isotopically rare species, especially C$^{18}$O, is also dominated by the envelopes, but to a lesser extent than $^{12}$CO, as the reduced
self-shielding of the rarer species in the UV-dominated transition
regions leads to isotope-selective dissociation and
fractionation.  The lower line optical depth in these
transitions probes the depth of the entire cloud, not just the surface
layers seen in low-J $^{12}$CO. 

Far-infrared [\ion{O}{1}] $^3P_0 \rightarrow {^3P_1}$ and
submillimeter HCO$^+$ 
line emission is only efficiently produced at higher
densities, so it doesn't significantly contribute to the line emission in the
diffuse envelope, but instead stems from the dense clumps.  The
HCO$^+ (4 \rightarrow 3)$ and $(3 \rightarrow 2)$ emission is readily
excited at low temperatures, and therefore stems from the cold, shielded,
molecular clump interiors as well as the UV-illuminated PDR surfaces
of clumps.   
The relatively high 228K excitation
energy of the [\ion{O}{1}] $^3P_0 \rightarrow {^3P_1}$
transition limits the vast majority of the intense [\ion{O}{1}]
emission to the warm surfaces of the clumps.  
However, if atomic oxygen is not entirely bound up in molecules like
CO, H$_2$O, or depleted onto dust grains in the shielded cores of
dense clouds, it can emit at detectable levels even at relatively cold
temperatures (T$_k=10-20$ K) at high densities ($n\sim10^6\ {\rm
cm}^{-3}$).  Indeed, it now seems certain that O$_2$ is not a significant
repository of elemental oxygen in molecular clouds (cf. Goldsmith et
al. 2000, and references therein); however, the nature of oxygen
depletion in molecular clouds is still debated enthusiastically.

A narrow range of density contrast is required to simultaneously account
for the modest strength of the 63 \micron\ [\ion{O}{1}] and submillimeter
HCO$^+$ emission.  At the higher densities ($n\sim10^6\ {\rm
cm}^{-3}$) needed to stimulate significant HCO$^+\ (4 \rightarrow 3)$ emission,
[\ion{O}{1}] emission can become very efficient and rapidly overwhelms
the 158 \micron\ [\ion{C}{2}] intensity for normal interstellar
conditions, contrary to ISO-LWS observations of $\rho$ Oph where
[\ion{C}{2}] emission is dominant (Liseau et al. 1999; Kaufman et al. 1999). 
However, clumps with nonuniform (power law) density, $n(r)$, are
theoretically expected and naturally satisfy both excitation conditions. 
As such, submillimeter HCO$^+$ emission would then originate primarily
from the dense cores of the clumps ($n \ga 10^6\ {\rm
cm}^{-3}$), whereas the [\ion{O}{1}] emission would stem from the
warmer ``surfaces'' of the clumps ($n \approx 10^5\ {\rm cm}^{-3}$).

\subsection{Other ISO positions}
\label{otheriso}

\begin{figure}[!ht]
\epsscale{0.8} \plotone{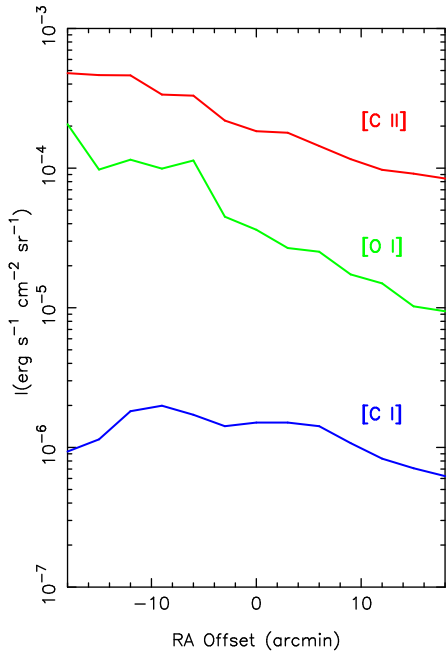}
\caption{\label{slice} Intensities of 158\micron\ [\ion{C}{2}],
63\micron\ [\ion{O}{1}] and 609\micron\ [\ion{C}{1}] line emission
along the ISO east-west cross-section of the $\rho$ Oph cloud}
\end{figure}

Our large scale maps of [\ion{C}{1}] overlap with all 13 pointings of
ISO-LWS from Liseau et al. (1999).  We can therefore perform limited
(homogenous) PDR modeling toward a 1-dimensional east-west cross
section of the $\rho$ Oph cloud.  In Figure~\ref{slice}, we plot the
cross-sectional intensities of 158\micron\ [\ion{C}{2}], 63\micron\
[\ion{O}{1}] and 609\micron\ [\ion{C}{1}] line emission.  As described
in Liseau et al. (1999), the [\ion{C}{2}] line emission decreases as projected
distance from HD~147889 increases, tracing the incident
radiation field upon the $\rho$ Oph PDR.  With a high critical
density and excitation energy, the [\ion{O}{1}] flux traces both
density and radiation field intensity.  [\ion{C}{1}] line emission
however shows a different spatial relation, 
instead following the cold gas column density as traced by C$^{18}$O
(for comparison, see Wilking \& Lada 1983); the two maxima in the
[\ion{C}{1}] distribution occur as the cross-sectional slice skirts
the edges of the $\rho$~Oph~A and $\rho$~Oph~B2 cloud cores.

As in \S~\ref{homopdr}, we use the ISO-LWS integrated
intensities of 158~\micron\ [\ion{C}{2}] and 63~\micron\ [\ion{O}{1}]
line emission to estimate $I_{\rm UV}$ and $n$.  The thickness of
the PDR model is decreased with increasing projected distance from
HD~147889 in order to correctly
recover the correct [\ion{C}{1}] line intensities.
Model results for four representative
points along the east-west cross section are presented in
Table~\ref{otherpos}. 

\begin{deluxetable*}{cccccc}
\tablecaption{\label{otherpos}PDR Modeling of a $\rho$ Oph cross-section;
$n=10^4$ cm$^{-3}$}
\tablewidth{0pt}
\tablehead{
\colhead{RA Offset\tablenotemark{(a)}} &   
\colhead{$I_{\rm UV}$} & 
\colhead{$T_{\rm PDR}$} &
\colhead{N(H$_2$)} & 
\colhead{63\micron\ [\ion{O}{1}]/} &
\colhead{609\micron\ [\ion{C}{1}]}\\
\colhead{(from WL1)} & & \colhead{(K)} & \colhead{(cm$^{-2}$)} &
\colhead{158\micron\ [\ion{C}{2}]}& \colhead{(erg s$^{-1}$ cm$^{-2}$ sr$^{-1}$)}
}
\tablecolumns{6}
\startdata
-18\arcmin=EW1 & 90 & 200 & $6\times 10^{21}$ & 0.4 &
$1\times 10^{-6}$ \\

-6\arcmin=EW5 & 30 & 100 & $1\times 10^{22}$ & 0.5 &
$2\times 10^{-6}$ \\

3\arcmin=EW8 & 20 & 100 & $1\times 10^{22}$ & 0.2 &
$1.5\times 10^{-6}$ \\

18\arcmin=EW13 & 10 & 50 & $4\times 10^{21}$ & 0.1 &
$8\times 10^{-7}$ \\
\enddata
\tablenotetext{(a)}{``EW'' positions correspond to Liseau et al. (1999)}
\end{deluxetable*}

\subsection{Directly Probing the Western H$_2$ PDR Front}
\label{h2}

Mapping across the western PDR front in the infrared emission lines of H$_2$ can clarify the cloud geometry at its ``true surface''. 
Boulanger
et al. (1999) presents ISO H$_2$ emission line spectra from 
the western PDR front of the $\rho$ Oph cloud in the 0-0 S(0) through
S(7), and 1-0 S(1) transitions.  The H$_2$ emission is observed to be
coincident with both PAH emission in the thermal IR and the
filamentary dust structure seen with ISOCAM at 7 and 15 \micron.  The
non-thermal population of high H$_2$ rotational levels coupled with
photoilluminated PAH emission suggests that UV fluorescence, rather
than thermal shocks, are likely to be responsible for the H$_2$
excitation.  Fluorescing H$_2$ emission directly
probes the incident UV field, PDR density, and geometry of the
\ion{H}{1}/H$_2$ transition region at A$_{\rm v} \la 1$ with high
angular resolution. 

Spectroscopy of fluorescent H$_2$ emission can also elucidate the
location of the H$_2$ dissociation region relative to the rest of the
molecular cloud.  For example, the 
H$_{2}$ 1-0 S(0,1) and  1-0 Q(2,3) line pairs arise from the same
upper states.  As these are both intrinsically weak electric-quadrupole
transitions, the effects of self-absorption can be safely neglected.
Thus the relative intensities of these two lines are 
functions only of their relative spontaneous radiative transition
probabilities and photon energies.  For example:

\begin{equation}
\frac{I_{\rm Q(3)}}{I_{\rm S(1)}}=\frac{A_{\rm Q(3)}}{A_{\rm S(1)}} \frac{\lambda_{\rm S(1)}}{\lambda_{\rm Q(3)}}
\end{equation}

\noindent for the quadrupole H$_{2}$ A-values computed most recently by
Wolniewicz, Simbotin \& Dalgarno (1998). 

In Table~{\ref{H2ext}}, line ratios are tabulated for visual extinction
(A$_{\rm v}$) from 0 to 40.  Such measurements, combined with the
intensity of the emission, would not only constrain the physical
conditions in which H$_2$ is dissociated, but also would clarify whether the
UV-illuminated surface that is fluorescing in H$_2$ is on the front or back
side of the $\rho$ Oph cloud.  

\begin{deluxetable}{ccc} 
\tablewidth{0pt}
\tablecaption{\label{H2ext} H$_2$ $v=1\rightarrow0$ line ratios as a function of increasing extinction}
\tablehead{
\colhead{A$_{\rm v}$} &
\colhead{I$_{\rm Q(3)}$/I$_{\rm S(1)}$} &
\colhead{I$_{\rm S(1)}$ \tablenotemark{(a)}}
}
\startdata
0 & 0.703 & $3.5 \times 10^{-6}$ \\
5 & 0.785 & $2.0 \times 10^{-6}$ \\
10 & 0.875 & $1.1 \times 10^{-6}$ \\
20 & 1.09 & $3.7 \times 10^{-7}$ \\
30 & 1.36 & $1.2 \times 10^{-7}$ \\
40 & 1.70 & $3.9 \times 10^{-8}$ \\
\enddata
\tablenotetext{(a)}{Extinguished intensity of 1-0 S(1), in ${\rm erg}\ {\rm s}^{-1}\ {\rm cm}^{-2}\ {\rm sr}^{-1}$}
\end{deluxetable}

\begin{figure}[!h]
\epsscale{0.9} \plotone{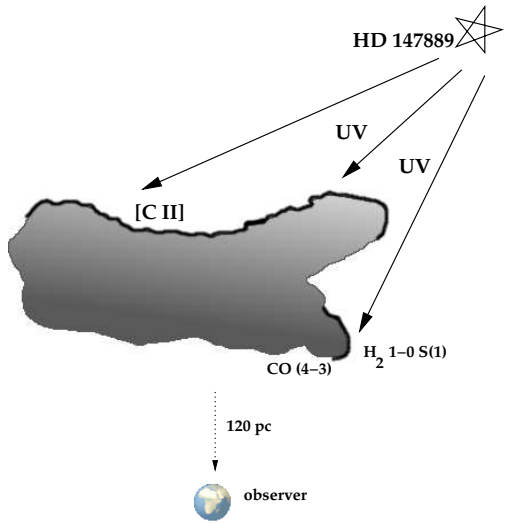}
\caption{\label{pdrpic} Schematic diagram of the $\rho$\ Oph dark cloud,
with PDR surfaces highlighted in black.  The concave back side of the
cloud, per Liseau et al. (1999), bears the bulk of the UV
radiation field from HD~147889.  However, if the extinction toward
the H$_2$ emission (Boulanger et al. 1999) is low, the ``filament''
PDR on the western edge of the dense molecular cloud likely represents
a foreground PDR region that is not shadowed from HD~147889 by the
rest of the cloud.  The depicted dark cloud is also surrounded by
a diffuse ``halo'' of gas and dust. 
}
\end{figure}

For example, at $I_{\rm UV} = 50$  and $n \sim
10^4\ {\rm cm}^{-3}$, the expected H$_2$ 1-0 S(1) line intensity is
modest, about $3.5 \times 
10^{-6}\ {\rm erg}\ {\rm s}^{-1}\ {\rm cm}^{-2}\ {\rm sr}^{-1}$.  If
the observed 1-0 S(1) emission (from Boulanger et al. 1999) originates
from the back side of the cloud, as does the rest of the [\ion{C}{2}]
emission, the near-IR H$_2$ emission would have to pass through a mean
column density loosely equivalent to A$_{\rm v} \sim 25$, assuming a
mean reference value of the gas/dust extinction ratio of $1.6 \times
10^{21}\ {\rm cm}^{-2}\ {\rm mag}^{-1}$ measured in diffuse
interstellar clouds.  This translates to a mean extinction of $\sim 3$
magnitudes at 2.12 \micron, interpolating the Rieke \& Lebofsky (1985)
infrared extinction curve.  This would extinguish the H$_2$ 1-0 S(1) line
flux by a factor of $\sim 17$, to $2.1 \times 10^{-7} \ {\rm erg}\
{\rm s}^{-1}\ {\rm cm}^{-2}\ {\rm sr}^{-1}$.  Noting this faintness,
and pending photometric calibration of the H$_2$ 1-0 S(1) images of
Boulanger et al. (1999), we predict that the H$_2$ emission instead
comes from a region of molecular material on the {\it front} side of
the cloud that is not shielded from HD~147889 and therefore is subject
to photodissociation.  A schematic view of the $\rho$~Oph PDR might
then look like Figure~\ref{pdrpic}.

Sensitive, high resolution near-infrared long slit spectroscopy of H$_2$ in
PDRs provides a natural complement to the kinds of long wavelength PDR
diagnostics descibed in this paper, and are worth renewed effort with
numerous far-infrared and submillimeter facilities soon to become
available (such as Herschel \& SOFIA).

\section{Summary}

In this paper, we have attempted to exploit the complementary
information afforded by large-scale and high-resolution submillimeter
spectroscopic observations made with small and large antennas. 

We present large scale maps ($40\arcmin \times 40\arcmin$),
obtained with AST/RO,
in CO$(4\rightarrow3)$ and [\ion{C}{1}]$(^3P_1\rightarrow{^3P_0})$
centered near the infrared source WL1 in the $\rho$ Ophiuchi molecular
cloud.  [\ion{C}{1}]$(^3P_1\rightarrow{^3P_0})$ 
emission is extended on
large scales, resembles the C$^{18}$O column density map presented by
Wilking and Lada (1983), and generally follows the edges of
warm dust emission expected for the illuminated surface of the
molecular cloud.  The optically-thick CO $(4 \rightarrow 3)$ emission
traces the CO ``surface'' of molecular clouds and is strongest
(hottest) at the western boundary of the molecular cloud, adjacent to
exciting B2V star HD~147889.  Adjacent ro-vibrational H$_2$ emission,
cospatial with a western filamentary structure in the hot dust
continuum at 7 \& 15 \micron\ and in PAH emission, provides evidence
for a new foreground PDR.  This is in contrast to the rest of the cloud, which
appears to be illuminated on the far side by HD~147889.
Additional H$_2$ spectroscopy is needed to verify these suppositions.

We also present smaller scale maps 
($\sim{5\arcmin \times 5\arcmin}$), taken with the HHT, in 
C$^{18}$O$(2 \rightarrow 1)$, C$^{18}$O$(3 \rightarrow 2)$,
$^{13}$CO$(2 \rightarrow 1)$, CO$(2 \rightarrow 1)$, CS$(5 \rightarrow
4)$, HCO$^+(3\rightarrow 2)$ and HCO$^+(4\rightarrow 3)$
toward a single region (EW4) in the AST/RO map, chosen to overlap
with [\ion{C}{2}] and [\ion{O}{1}] observations made by  Liseau et
al. (1999) with ISO-LWS.  The C$^{18}$O structure of the $\rho$~Oph~A
core is directly comparable to recent submillimeter continuum
emission, however HCO$^+$ line emission is concentrated in a ring-like
structure.  Modeling of the HCO$^+$ ring in comparison to CO suggests
that the condensations are due to both a higher HCO$^+$ abundance and
a dominant dense gas component with $n\sim10^6\ {\rm cm}^{-3}$.  A
slightly enhanced ionization rate contributes to the elevated HCO$^+$
abundance in this region.  
Although it also stems from dense gas, the emission from
CS follows a totally different morphology.
The $\rho$~Oph~A cloud core contributes
to a plateau of CS emission, but the actual CS peak is coincident
with the relatively minor $\rho$~Oph~A2 core and outflow instead.  Shock and
outflow chemistry is suspected to contribute to a higher CS abundance
in this region.  

We use the emission lines presented in this paper to infer 
column densities of all optically thin species toward ISO-EW4, and
three other positions examined by ISO-LWS. 
We use the indirect determinations of 
H$_2$ column density from C$^{18}$O and $^{13}$CO with PDR models 
to constrain the
physical conditions in which [\ion{O}{1}] and [\ion{C}{2}] line
emission is observed (Liseau et al. 1999).  We model the physical
structure of the $\rho$ Oph PDR in the traditional homogeneous,
plane-parallel formalism, and obtain self-consistent fits to most
available observations.  However, the luminous and clumpy
submillimeter HCO$^+$ and infrared [\ion{O}{1}] emission directly
implies dense substructure.  We adopt a simple two-component PDR
model, featuring a diffuse gaseous  
envelope in which dense clumps reside, to account for the HCO$^+$ and
[\ion{O}{1}] 
emission.  In this picture of the physical structure of the PDR, we
are able to attribute emission from different species to different
physical locations within the $\rho$ Oph PDR. 
We find that [\ion{C}{2}] and [\ion{C}{1}] 
emission predominantly arises from the lower density envelopes.  We
find little evidence for non-equilibrium chemistry providing a
``carbon-rich'' region as per Kamegai et al. (2003), finding instead
that the atomic carbon distribution follows the C$^{18}$O ridge
through dust continuum peaks E and F.   We find that like
[\ion{C}{1}], low-J CO emission stems predominantly from the
envelopes, and high-J emission should stem from the denser clumps.  
Submillimeter HCO$^+$ and infrared [\ion{O}{1}] emission stems from
the clumps and clump surfaces, respectively, and indicate
clump surface temperatures of $50-200$ K, an ultraviolet radiation
field with I$_{\rm UV} \sim 10-90$ (G$_0 \sim 10^2$),
densities of 10$^5$ to 10$^6$ cm$^{-3}$, and interior temperatures of
$\leq20$K.  

This study illustrates the use of far-infrared, submillimeter, and
infrared spectroscopy for dissecting the complicated structure of
clouds and star forming regions.  Yet, 
numerous new advances in detector technology will soon enhance our
understanding more dramatically, putting 
the evolution of clouds and stars in a more global, Galactic context.
The advent of 
large-format infrared detector arrays will make broad and narrow band
observations of hot dust and gas possible over entire clouds, as well
as providing photometry for detailed extinction maps.  Although
integrated heterodyne detector arrays at submillimeter wavelengths are
in their infancy, rapid progress is expected to revolutionize the
radio universe as arrays have already done for the infrared.  Finally,
new airborne and space-based platforms such as the Spitzer Space
Telescope, Herschel, SOFIA, and SAFIR will open up the far infrared,
opening new windows into the chemical history of carbon, nitrogen,
oxygen and water in the Galaxy.  

\acknowledgements

This work is based in part on measurements made with
the Heinrich Hertz Telescope, which is 
operated by the Submillimeter Telescope
Observatory on behalf of Steward Observatory 
and the Max-Planck-Institut f\"ur Radioastronomie.

It is a special pleasure to thank the scientific and technical staff of the
Heinrich Hertz Submillimeter Telescope 
Observatory (HHT) for observational documentation and assistance at the
telescopes.  This research was supported in part by the National Science 
Foundation under a cooperative agreement with the Center for 
Astrophysical Research in Antarctica (CARA), grant~NSF~OPP~89-20223.
CARA is a National Science Foundation Science and Technology Center.

\enlargethispage{1.0cm}


\begin{thebibliography}{}

\bibitem[Abergel et al.(1996)]{abe96} 
Abergel, A.\ et al.\ 1996, \aap, 315, L329

\bibitem[1993]{and93}
Andr\'e, P., Ward-Thompson, D.\, \& Barsony, M.\ 1993, \apj, 406, 122 

\bibitem[Ashby et al.(2000)]{2000ApJ...539L.119A} Ashby, M.~L.~N., et al.\ 
2000, \apjl, 539, L119 

\bibitem[1987]{bal87}
Bally, J., Stark, A. A., Wilson, R. W., \& Langer, W. D. 1987, \apj, 312, L45

\bibitem[Black \& van Dishoeck(1987)]{BvD87} Black, J.\ H.\
\& van Dishoeck, E.\ F.\ 1987, \apj, 322, 412

\bibitem[Blitz \& Williams(1997)]{1997ApJ...488L.145B} Blitz, L.~\&
Williams, J.~P.\ 1997, \apjl, 488, L145

\bibitem[Blum \& Pradhan(1992)]{1992ApJS...80..425B} Blum, R.~D.~\& Pradhan, A.~K.\ 1992, \apjs, 80, 425. 

\bibitem [1990]{boi90} 
Boiss\'e, P. 1990, A\&A, 228, 483

\bibitem[1999]{bou99}
Boulanger, F., Abergel, A., Falgarone, E., Habart, E., Pineau Des
For{\^e}ts, G., \& Verstraete, L.\ 1999, H$_2$ in Space, meeting held in
Paris, France, September 28th - October 1st, 1999.\ Eds.: F.\ Combes,
G.\ Pineau des For{\^e}ts.\ Cambridge University Press, Astrophysics
Series, E34 

\bibitem [1990]{bur90} 
Burton, M. G., Hollenbach, D. J., \& Tielens, A. G. G. M. 1990,
ApJ, 365, 620

\bibitem[1986]{car86}
Cardelli, J.\ A.\ \& Wallerstein, G.\ 1986, \apj, 302, 492 

\bibitem[Carrasco, Strom, \& Strom(1973)]{1973ApJ...182...95C}
Carrasco, L., Strom, S.~E., \& Strom, K.~M.\ 1973, \apj, 182, 95


\bibitem[1998]{cec98}
Ceccarelli, C., Caux, E., White, G. J., Molinari, S., Furniss, I., Liseau, R.,
Nisini, B., Saraceno, P., Spinoglio, L., \& Wolfire, M. 1998, A\&A, 331, 372

\bibitem[1985]{cru85}
Crutcher, R.\ M.\ \& Chu, Y.\ 1985, \apj, 290, 251 

\bibitem[1980]{dej90}
de Jong, T., Boland, W., \& Dalgarno, A.\ 1980, \aap, 91, 68 

\bibitem[Draine \& Bertoldi(1996)]{Dra96} Draine, B.\ T.\ \&
Bertoldi, F.\ 1996, \apj, 468, 269

\bibitem[Engargiola, Zmuidzinas, \& Lo(1994)]{eng94}
Engargiola, G., Zmuidzinas, J., \& Lo, K.\ Y.\ 1994, Review of Scientific
Instruments, 65, 1833

\bibitem[Flower(1990)]{1990MNRAS.242P...1F} Flower, D.~R.\ 1990, \mnras, 
242, 1P. 

\bibitem[Flower \& Pineau-Des-Forets(1990)]{flo90} Flower, 
D.~R.~\& Pineau-Des-Forets, G.\ 1990, \mnras, 247, 500. 


\bibitem[1989]{fre89}
Frerking, M. A., Keene, J., Blake, G., \& Phillips, T. G.  1989, ApJ, 344, 311

\bibitem[Goldsmith et al.(2000)]{gol00} Goldsmith, P.\ F.\ et
al.\ 2000, \apjl, 539, L123

\bibitem[Hamaguchi, Corcoran, \& Imanishi(2003)]{2003PASJ...55..981H} 
Hamaguchi, K., Corcoran, M.~F., \& Imanishi, K.\ 2003, \pasj, 55, 981 

\bibitem [1991] {hol91}
Hollenbach, D. J., Takahashi, T., \& Tielens, A. G. G. M. 1991,
ApJ, 377, 192

\bibitem [1991] {how91}
Howe, J. E., Jaffe, D. T., Genzel, R., \& Stacey, G. J. 1991,
ApJ, 373, 158

\bibitem[Imanishi, Tsujimoto, \& Koyama(2002)]{2002ApJ...572..300I} 
Imanishi, K., Tsujimoto, M., \& Koyama, K.\ 2002, \apj, 572, 300 

\bibitem[Irvine, Goldsmith, \& Hjalmarson(1987)]{irv87}
Irvine, W.\ M., Goldsmith, P.\ F., \& Hjalmarson, A.\ 1987, ASSL Vol.\ 134:
Interstellar Processes, 561

\bibitem[Jaquet, Staemmler, Smith, \& Flower(1992)]{1992JPhB...25..285J} 
Jaquet, R., Staemmler, V., Smith, M.~D., \& Flower, D.~R.\ 1992, Journal of 
Physics B Atomic Molecular Physics, 25, 285. 

\bibitem[Johnstone et al.(2000)]{2000ApJ...545..327J} Johnstone, D.,
Wilson, C.~D., Moriarty-Schieven, G., Joncas, G., Smith, G.,
Gregersen, E., \& Fich, M.\ 2000, \apj, 545, 327

\bibitem[Kaufman, Wolfire, Hollenbach, \&
Luhman(1999)]{kau99} Kaufman, M.\ J., Wolfire, M.\ G.,
Hollenbach, D.\ J., \& Luhman, M.\ L.\ 1999, \apj, 527, 795

\bibitem[1985]{kee85}
Keene, J., Blake, G., Phillips, T. G., Huggins, P. J., \& Beichman,
C. A.\ 1985, ApJ, 299, 967

\bibitem[1987]{kee87}
Keene, J.  1987, SETI Conference {\it{Carbon in the Galaxy}}

\bibitem [1994] {kos94}
K\"oster, B., St\"orzer, H., Stutzki, J., \& Sternberg, A. 1994,
A\&A, 284, 545

\bibitem[Langer(1985)]{1985prpl.conf..650L} Langer, W.~D.\ 1985, Protostars 
and Planets II, 650 

\bibitem [1993] {leb93}
le Bourlot, J., Pineau des For\^ets, G., Roueff, E., \&
Flower, D. R. 1993, A\&A, 267, 233

\bibitem [1999] {lis99}
Liseau, R., White, G., Larsson, B., Sidher, S., Olofsson, G.,
Kaas, A., Nordh, L., Caux, E., Lorenzetti, D., Molinari, S., Nisini, B., \&
Sibille, F. 1999, A\&A, 344, 342

\bibitem[Maloney, Hollenbach, \& Tielens(1996)]{1996ApJ...466..561M} 
Maloney, P.~R., Hollenbach, D.~J., \& Tielens, A.~G.~G.~M.\ 1996, \apj, 
466, 561 

\bibitem[Mauersberger et al.(1992)]{mau92} Mauersberger, R.,
Wilson, T.\ L., Mezger, P.\ G., Gaume, R., \& Johnston, K.\ J.\ 1992, \aap,
256, 640

\bibitem [1993] {mei93}
Meixner, M., \& Tielens, A. G. G. M. 1993, ApJ, 405, 216

\bibitem[Mitchell(1984)]{mit84} Mitchell, G.~F.\ 1984, \apj, 287, 665 

\bibitem[Monteiro(1985)]{1985MNRAS.214..419M} Monteiro, T.~S.\ 1985, \mnras, 214, 419

\bibitem[Motte, Andr\'e, \& Neri(1998)]{1998A&A...336..150M} Motte, F.,
Andr\'e, P., \& Neri, R.\ 1998, \aap, 336, 150

\bibitem[Pequignot(1990)]{1990A&A...231..499P} Pequignot, D.\ 1990, \aap, 
231, 499. 

\bibitem[1992]{pin92}
Pineau Des For\^ets, G., Roueff, E., \& Flower, D.\ R.\ 1992, \mnras, 258, 45P 

\bibitem[1999]{plu99}
Plume, R., Jaffe, D.\ T., Tatematsu, K., Evans, N.\ J., \& Keene, J.\ 1999, \apj, 512, 768 

\bibitem[1985]{rie95}
Rieke, G.\ H.\ \& Lebofsky, M.\ J.\ 1985, \apj, 288, 618 

\bibitem[Schieder, Tolls, \& Winnewisser(1989)]{sch89}
Schieder, R., Tolls, V., \& Winnewisser, G.\ 1989, Experimental Astronomy,
1, 101

\bibitem[Schinke et al.(1985)]{1985ApJ...299..939S} Schinke, R.,
Engel, V., Buck, U., Meyer, H., \& Diercksen, G.~H.~F.\ 1985, \apj,
299, 939

\bibitem[Schr\"oder et al.(1991)]{sch91} Schr\"oder, K., Staemmler, V., Smith, M. D., Flower, D. R., \& Jaquet,  R. 1991, J. Phys. B, 24, 2487  

\bibitem[1977]{shu77}
Shu, F.\ H.\ 1977, \apj, 214, 488 

\bibitem[Shu, Adams, \& Lizano(1987)]{Shu87} Shu, F.\ H.,
Adams, F.\ C., \& Lizano, S.\ 1987, \araa, 25, 23

\bibitem[Snell et al.(2000)]{2000ApJ...539L.101S} Snell, R.~L., et al.\ 
2000, \apjl, 539, L101 

\bibitem [1996] {spa96}
Spaans, M. 1996, A\&A, 307, 271

\bibitem[Stark et al.(2001)]{sta01} Stark, A.\ A.\ et al.\
2001, \pasp, 113, 567

\bibitem [1988] {stu88}
Stutzki, J., Stacey, G. J., Genzel, R., Harris, A. I.,
Jaffe, D. T., \& Lugten, J. B. 1988, ApJ, 332, 379

\bibitem[Tauber \& Goldsmith(1990)]{1990ApJ...356L..63T} Tauber,
J.~A.~\& Goldsmith, P.~F.\ 1990, \apjl, 356, L63. 

\bibitem[1985]{tie85a}
Tielens, A. G. G. M. \& Hollenbach, D.\ 1985, ApJ, 291, 722

\bibitem[2002]{wal02}
Walker, C. K., Kooi, J., \& Jacobs, K.\ 2002, in preparation.

\bibitem[1983]{wil83}
Wilking, B. A. \& Lada C. J. 1983, ApJ, 274, 698

\bibitem[Wilson et al.(1999)]{1999ApJ...513L.139W} Wilson, C.~D.~et
al.\ 1999, \apjl, 513, L139

\bibitem[1999]{wil94}
Wilson, T.\ L.\ \& Rood, R.\ 1994, \araa, 32, 191

\bibitem[1984]{vand84}
van Dishoeck, E.\ F.\ \& de Zeeuw, T.\ 1984, \mnras, 206, 383 

\bibitem[van Dishoeck \& Black(1988)]{vdb88} van Dishoeck,
E.\ F.\ \& Black, J.\ H.\ 1988, \apj, 334, 771

\bibitem[1998]{wol98}
Wolniewicz, L., Simbotin, I., \& Dalgarno, A.\ 1998, \apjs, 115, 293 

\bibitem[1999]{zmu99}
Zmuidzinas, J. \& LeDuc, H. G. 1999, IEEE Trans. Microwave Theory Tech., 
40, 1797

\end{thebibliography}
\end{document}